\documentclass[runningheads]{llncs}
\usepackage{graphicx}
\usepackage{caption}
\usepackage{amsmath}
\usepackage{amssymb}

\DeclareMathOperator*{\argmin}{argmin}
\begin{document}
\title{The Reincarnation of Grille Cipher: A Generative Approach}
%
%
\author{Jia Liu\inst{*} \and
Yan Ke \and
Yu Lei \and
Jun Li \and
Yaojie Wang\and
Yiliang Han \and
Minqing Zhang \and 
Xiaoyuan Yang  
}

%

\institute{
Key Laboratory of Network and Information Security, School of Cryptology Engineering,  Engineering University of People Armed Police Force, Xi'an 710086, China\\
\email{liujia1022@gmail.com}}

\maketitle                 
\begin{abstract}
In order to keep the data secret, various techniques have been implemented to encrypt and decrypt the secret data. Cryptography is committed to the security of content, i.e. it cannot be restored with a given ciphertext. Steganography is to hiding the existence of a communication channel within a stego. However, it has been difficult to construct a cipher (cypher) that simultaneously satisfy both channel and content security for secure communication. Inspired by the Cardan grille, this paper presents a new generative framework for grille cipher. A digital cardan grille is used for message encryption and decryption. The ciphertext is directly sampled by a powerful generator without an explicit cover. Message loss and prior loss are proposed for penalizing message extraction error and unrealistic ciphertext. Jensen-Shannon Divergence is introduced as new criteria for channel security. A simple practical data-driven grille cipher is proposed using semantic image inpainting and generative adversarial network. Experimental results demonstrate the promising of the proposed method.

\keywords{Grille cipher  \and Cryptography \and Steganography \and Image inpainting \and Generative adversarial network.}
\end{abstract}

\section{Introduction}
\begin{figure}
\includegraphics[width=\textwidth]{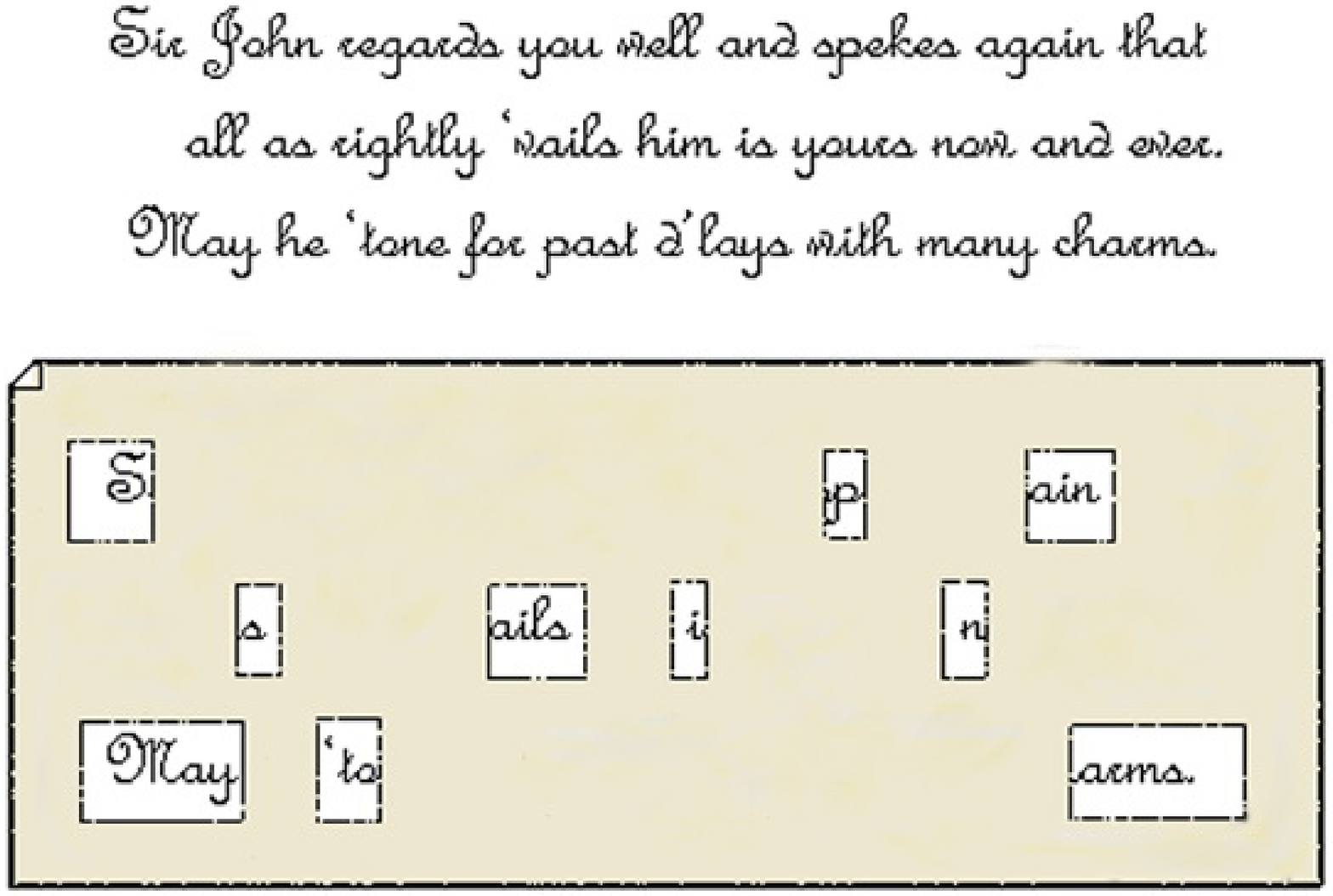}
\caption{A Cardan grille has no fixed pattern \cite{ref_wikipedia_cardangrille}.} \label{fig_cardangrille}
\end{figure}

In the history of cryptography, a grille cipher was a technique for encrypting a plaintext by writing it onto a sheet of paper through a pierced sheet (of paper or cardboard or similar) \cite{ref_wikipedia_grille}. The earliest known description is due to the polymath Girolamo Cardano, known in French as $J\acute{e}r\hat{o}me\quad Cardan$ in 1550. His proposal was for a rectangular stencil allowing single letters, or words to be written, then later read, through its various apertures. The written fragments of the plaintext could be further disguised by filling the gaps between the fragments with anodyne words or letters, as shown in Fig.~\ref{fig_cardangrille}\cite{ref_wikipedia_cardangrille}. This variant is also an example of steganography, as are many of the grille ciphers.

The Cardan grille was invented as a method of secret writing. The word cryptography became the more familiar term for secret communications from the middle of the 17th century. Earlier, the word steganography was common. The other general term for secret writing was cipher.  Sir Francis Bacon gave three fundamental conditions for ciphers. Paraphrased, these are\cite{ref_wikipedia_grille}:

1. A cipher method should not be difficult to use;

2. It should not be possible for others to recover the plaintext;

3. In some cases, the presence of messages should not be suspected.

It is difficult to fulfil all three conditions simultaneously at that time. There is a modern distinction between cryptography and steganography. Condition 3 applies to steganography. Bacon meant that a cipher message should, in some cases, not appear to be a cipher at all. The original Cardan Grille met that aim. However, the original method is slow and requires literary skill. Above all, any physical cipher device is subject to loss, theft and seizure; so to lose one grille is to lose all secret correspondence constructed with that grille. Variations on the Cardano original, however, were not intended to fulfill condition 3 and generally failed to meet condition 2 as well.

Unfortunately, design a cipher scheme satisfying all these constrains is still a difficult problem so far. Cryptography try to converse a readable state to ciphertext which is apparent nonsense, and steganography try to make the ciphertext looks normal so that the presence of messages should not be suspected. The advantage of steganography over cryptography alone is that the intended secret message does not attract attention to itself as an object of scrutiny. However, for a long time, the difficulty of constructing the \emph{ordinary letter} made steganography nearly going to be information modification. These traditional steganography methods such as \cite{ref_article_Sallee} and \cite{ref_article_Filler}, by modifying the cover data to hide the information, did not strictly satisfy the condition 3, had to struggle against the steganalysis technique\cite{ref_article_Fridrich}.

In this paper, we come back to the road of Cardan, a new generative cipher framework is proposed. We consider the new cipher as a constrained generation problem and take advantage of the recent advance in generative modeling. After a deep generative model,i.e., in our case an adversarial net work, is trained, we search for an encoding that is ``closest''to the perfect ciphertext in latent ciphertext space. The encoding is then used to construct a ciphertext using generator. We define  ``closest'' by a message loss to penalizes message extraction error, and a prior loss to penalizes unrealistic samples. A symmetric key called digital Cardan grille is used for both encryption and decryption. The realistic sample generation and key space, theoretically, guaranteed to meet the above three fundamental conditions. A practical cipher is proposed using image inpainting which is a particular application in image synthesis. Firstly, the corrupted image is taken as the cover, and the secret information is written to the area that needs to be remain unchanged with digital Cardan grille. The semantic image completion is realized by using the generative adversarial network with message loss and prior loss. The secret message is hidden in the reconstructed image after completion. The experiments on the image database confirms the promising of such simple method.

\section{Related Work}
Considering the respective advantages and disadvantages of steganography and cryptography, it is naturally to get an idea that combining them would simultaneously takes the advantages of steganography and cryptography while avoid the respective defects. In earlier work, there have been a lot of works applying this idea and one may refer to \cite{ref_lncs_Sharp} and ~\cite{ref_article_Li}, but most of these methods do the encryption and hiding separately, they encrypt the secret information firstly and then hide them in the digital media.
There has been fundmental work on provably secure steganography,  Cachin~\cite{ref_lncs_Cachin} introduced an information-theoretic model for steganography. Hopper etc.~\cite{ref_lncs_Hopper} have given a theoretical framework for steganography based on computational security. Le \cite{ref_article_Le} presented ideas for improving the efficiency of scheme  and Backes etc.~\cite{ref_lncs_Backes} proposed a modification which makes the scheme secure against a more powerful active adversary. Ahn etc. ~\cite{ref_lncs_Ahn} provided a formal framework for public-key steganography and to prove that public-key steganography is possible. Song etc.~\cite{ref_article_Song} proposed a method doing encryption and hiding at the same time, organically combining steganography and cryptography. This protocol is based on the LSB matching method in steganography and Boolean functions used in cryptography.

In Fridrich's groundbreaking work of modern steganography \cite{ref_book_Fridrich}, steganographic channel is divided into three categories, cover selection, modification and synthesis. steganography focus on the conditions 3 as show in above. Cover selection method does not modify the cover image, thereby avoiding the threat of the existing steganalysis technology. This method cannot be applied to practical applications because of its low payload. Cover modification is the most studied method so far. In terms of KL divergence as a security measure, it can only achieve $\epsilon$-security or the perfect security for a certain explicit model. Cover synthesis seems more consistent with the earlier Cardan grille. However, about ten years ago, this method is only a theoretical conception, rather than a practical steganography, because it is difficult to obtain multiple samples. With the help of texture synthesis, \cite{ref_proc_Otori,ref_article_Otori} use the texture sample and a bunch of color points generated by secret messages to construct dense texture images. \cite{ref_article_Wu} improves the embedding capacity by proportional to the size of the stego texture image. Qian etc. \cite{ref_proc_Qian} propose a robust steganography based on texture synthesis. Xue etc. \cite{ref_article_Xu} use marbling, a unique texture synthesis method that allows users to deliver personalized messages with beautiful, decorative textures for hiding message. This kind of texture-based methods are based on the premise that the cover may not represent the content in real world which is counter-intuitive for steganography which objective is to maintain the nature content of the cover.

Fortunately, a data-based sampling technique, generative adversarial networks (GANs) \cite{ref_proc_Goodfellow} have become a new research hot spot in artificial intelligence. Recently, two types of designs have applied adversarial training to cryptographic and steganographic problems. Abadi \cite{ref_article_Abadi} used adversarial training to teach two neural networks to encrypt a short message that fools a discriminator. However, it is hard to offer an evaluation to show that the encryption scheme is computationally difficult to break.  Instead of relying on manual password analysis, PassGAN\cite{ref_article_Hitaj} uses a GAN to autonomously learn the distribution of real passwords from actual password leaks, and to generate high-quality password guesses. Adversarial training has also been applied to steganography. Volkhonskiy etc.\cite{ref_article_Volkhonskiy} first propose a new model for generating image-like containers based on Deep Convolutional Generative Adversarial Networks (DCGAN\cite{ref_article_Radford}). This approach allows to generate more setganalysis-secure message embedding using standard steganography algorithms,they do not measure performance against state-of-the-art steganographic techniques making it difficult to estimate the robustness of their scheme.Similar to \cite{ref_article_Volkhonskiy}, Shi etc.\cite{ref_article_Shi} introduce a new generative adversarial networks to improve convergence speed, the training stability the image quality. Similar to \cite{ref_article_Abadi}, Hayes~\cite{ref_article_Hayes} define a game between three parties, Alice, Bob and Eve, in order to simultaneously train both a steganographic algorithm and a steganalyzer. However£¬Alice is still trained to learn to produce a steganographic image by LSB which is a traditional cover modification method. Tang etc.\cite{ref_article_Tang} propose an automatic steganographic distortion learning framework using a generative adversarial network, which is composed of a steganographic generative subnetwork and a steganalytic discriminative subnetwork. However, most of these GAN-based steganographic schemes are still the cover modification techniques. These methods focus on the adversarial game between steganography and steganalysis while ignoring the core aim of the GAN is to build a powerful sampler.

Since GAN's biggest advantage is to generate samples, it is a intuitive idea to use GANs generate a semantic cipher from a message directly as the Cardan did. Some researcher made a preliminary attempt on this intuitive idea. Ke \cite{ref_article_Ke} proposed generative steganography method called GSK in which the secret messages are generated by a cover image using a generator rather than embedded into the cover, thus resulting in no modifications in the cover. Liu etc. \cite{ref_article_Liu} propose a method that using ACGANs \cite{ref_article_Odena} to classify the generated samples, and they make the class output information as the secret message. In \cite{ref_article_Jia}, the secret message is written to the corrupted area of image that needs to be filled, then the corrupted stego image is fed into a Generative Adversarial Network (GAN) for stego generation.

The proposed scheme in this paper can be considered as a steganographic scheme as well as a classical cryptographic scheme, so we call the proposed scheme a cipher scheme. Different from classical cryptography and cover modification steganography, ciphertext is sampled by a generator with constrains. In this paper, our main work is to take advantage of machine learning approach, a deep generative model by adverserial network, to implemente a classical cryptography(cardan grille). The content security of our proposed scheme is equivalent to classical Cardan grille cipher. To the best of our knowledge, we are the first to provide a formal framework for generative cipher and to prove that channel security is possible. In order to maintain consistency with modern stegonagraphy, we also use the term stego to denote the ciphertext in this paper. Our generative cipher can be also considered as a generative steganography. This paper has the following contributions:

1. We propose a practical data-driven framework call digital Cardan grille for cipher by learning a generator, ciphertext is directly sampled by a generator. This framework simplifies the design of cipher. Classical cipher design to a large extent be automated. It can also be applied to other media, such as text, video and other fields. This scheme is also a key-dependent steganographic scheme adhere to Kerckhoffs's principle.

2. A new criterion for channel security (steganography) is defined. A formal representation for cipher is proposed which figure out the connection between steganography and classical cryptography. We also give a toy example of cipher scheme to illustrate the relationship between classical cryptography and steganography.

3. Compared with texture synthesis methods, semantic image inpainting is used to ensure the logical rationality of cover contents. Compared with cover modification steganography, our method make the ciphertext distribution and real data distribution as close as possible, there is no specific original cover. Other image synthesis methods ~\cite{ref_articel_Huang} based on the generative model can be easily converted into a cipher scheme using this framework.

The remainder of this letter is organized as follows: We detail the formal representation of cipher in the following section. Section IV show how to construct a cardboard grille ciphers using constrained image generation by GANs. Experiment results are demonstrated in Section V. Section VI concludes this research and details our future work.

\section{Framework}

The Generative Cipher (steganography) Framework (CCF) of this paper is shown in Fig.~\ref{fig_generativeframework} as follows:

\begin{figure}
\includegraphics[width=\textwidth]{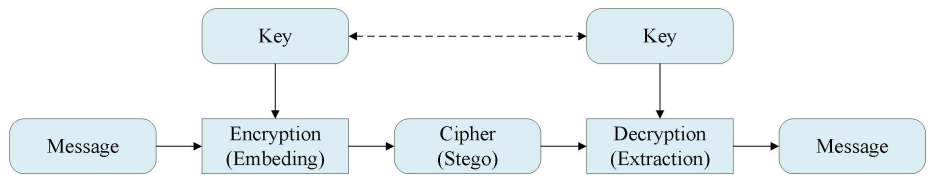}
\caption{Generative steganography framework.} \label{fig_generativeframework}
\end{figure}

\noindent In this scenario, the sender create a stego carrier from a generator with message directly. The encryption algorithm actually turns into a cipher generation process. The secret key shared by both parties ensures the security of the message, the natural real degree of stego determines the security of the communication channel.

In this paper we consider the notion of generative steganography against adversaries that do not attempt to disrupt the communication between Alice
and Bob (i.e., the goal of the adversary is only to detect whether steganography is being used and not to disrupt the communication between the participants). We show that secure cipher exists if any of several assumptions hold. Furthermore, we introduce a practical scheme that is secure under these assumptions. It is important to note that the shared key here is indispensable, in the case of  active attack defined in steganography, if the adversary can easily get the key, then the communication is not safe.

\subsection{Perfect Cipher Conditions}
In this section, we formalize cipher algorithm. Ideally the cipher scheme should satisfy the following three fundamental conditions which in line with the idea of Sir Francis Bacon. We call these perfect cipher conditions, or PCC:
\begin{equation}
c=E(m,k)
\end{equation}
\begin{equation}
m=D(c,k)
\end{equation}
\begin{equation}
p_{cipher} = p_{real}
\end{equation}
\noindent where original information $m$ is known as plaintext, $c$ denotes the encrypted form as ciphertext. $k$ is the symmetric key shared by two parties. E(.) in condition (1) is a encryption algorithm of cipher, and D(.) is the decryption operation. $p_{cipher}$ and $p_{real}$  denotes the distribution of ciphertext and real data. PCC (1) and (2) make the framework close to classical cryptography, which was effectively synonymous with encryption, the conversion of information from a readable state to apparent nonsense. PCC (3) guarantees the security of communication channel which is aim of steganography. Therefore, our goal for perfect cipher algorithm is to find a cryptography that satisfying steganography condition $p_{cipher} = p_{real}$.

A data-based sampling technique, generative adversarial network (GAN) is to estimate the potential distribution of existing data and generate new data samples from the same distribution. If we can learn a powerful generator which satisfy the PCC (3), together with a classical cryptography, a perfect cipher solution can be achieved to protect both content and channel of the secret information. In this paper, inspired by the intuitive idea, we use generator of GAN to sample a semantic ciphertext driven by a message directly. We call this generative cipher or generative stegonoraphgy. In generative cipher, we not only require the $p_{cipher} =p_{real}$, but also require the generator to satisfy the request for message extraction as shown in PCC (2). Therefore, we consider generative cipher as a constrained ciphertext generation problem and take advantage of the recent advances in generative modeling.

Interestingly, suppose all these conditions are satisfied, if every ciphertext sampled from generator is exactly as same as a real data sample. The generator can be seen as a way to select real samples from the world. This also means that the generator will be able to construct an infinite real sample database. You can imagine that the generator is a simulated digital camera, and each sampling is equivalent to taking a picture from the real world. What's even more amazing is that every seemingly normal picture contains a secret message.

\subsection{A Measure of Channel Security}
In mathematical statistics, the Kullback\text{-}Leibler divergence is a measure of how one probability distribution diverges from a second, expected probability distribution. Fridrich \cite{ref_book_Fridrich} introduces a formal information theoretic definition of security in steganography based on the Kullback\text{-}Leibler divergence between the distributions of cover and stego objects:
\begin{equation}
D_{KL}(p_{stego} \left| \right|p_{cover})=E_{x \sim p_{cover}}[\log \frac{p_{stego}}{p_{cover}}]=E_{x \sim p_{cover}}[\log p_{stego}-\log p_{cover}]
\end{equation}

\noindent where $p_{cover}$ and $p_{stego}$ are the distributions of cover and stego, respectively. However, KL divergence doesn't satisfy the symmetric and triangle inequality conditions, it cannot be strictly considered as a metric. The security of different steganography cannot be evaluated with this divergence. In this paper, a new measure of security for steganography is defined by the Jensen\text{-}Shannon divergence, which is based on the Kullback\text{-}Leibler divergence, with some notable (and useful) differences, including that it is symmetric and it is always a finite value. It is defined by:
\begin{equation}
D_{JS}(p_{stego} \left| \right|p_{cover})=\frac{1}{2}D(p_{stego} \left| \right|M)+\frac{1}{2}D(p_{cover} \left| \right|M)
\end{equation}
\begin{equation}
M=\frac{1}{2}(p_{stego}+p_{cover})
\end{equation}

The Jensen\text{-}Shannon divergence is bounded by 1 for two probability distributions, given that one uses the base 2 logarithm
\begin{equation}
0 \leq D_{JS}(p_{stego} \left| \right|p_{cover}) \leq 1
\end{equation}

\noindent when $p_{stego}$ = $P_{cover}$, Jensen\text{-}Shannon divergence is zero.

In generative steganography, we can use this metric to evaluate which generator is closer to the real data distribution. It means that we can sample a security stego from the best generator. In fact, the generator in generative adversarial network \cite{ref_proc_Goodfellow} is trained based on the Jensen\text{-}Shannon divergence, the adversarial game make the divergence between generator distribution $p_{g}$ and data distribution $p_{data}$ is gradually reduced with the increasing of adversarial iteration. In generative steganography, $p_{g}$, $p_{stego}$ and $p_{fake}$ have the similar meaning, $p_{cover}$ is $p_{data}$ . Since in our scheme, there is no explicit cover, we use $p_{data}$ instead of $p_{cover}$.

In \cite{ref_proc_Goodfellow}, Goodfellow etc. train discriminative model $D$ to maximize the probability of assigning the correct label to both training examples and samples from $G$. They simultaneously train G to minimize $log(1 - D(G(z)))$. In other words, $D$ and $G$ play the following two-player minimax game with value function $V (G; D)$:
\begin{equation}
\min_{G}\max_{D}V(D,G)=E_{x\sim p_{data}(x)}[\log D(x)]+E_{z\sim p_{z}(z)}[\log (1-D(G(z)))]
\end{equation}

\noindent The minimax game in Eq. 8 can be reformulated as:
\begin{equation}
C(G)=E_{x\sim p_{data}}[\log \frac{p_{data}(x)}{p_{data}(x)+p_{g}(x)}]+E_{x\sim p_{g}}[\log \frac{p_{g}(x)}{p_{data}(x)+p_{g}(x)}]
\end{equation}

\noindent The theorem and proposition are given with their proof in ~\cite{ref_proc_Goodfellow} for theoretical  proving the convergence of algorithm.
\begin{theorem}
The global minimum of the virtual training criterion C(G) is achieved if and only if
$p_{g} = p_{data}$. At that point, C(G) achieves the value  $-\log 4$.
\end{theorem}
\begin{equation}
C(G)=\max_{D}V(G,D)= -\log 4+D_{JS}(p_{data},p_{g})
\end{equation}
%
\begin{proposition}
If G and D have enough capacity, the discriminator D
is allowed to reach its optimum given G, and $p_{g}$ is updated so as to improve the criterion :
\begin{equation}
E_{x\sim p_{data}}[\log D_{G}^{*}(x)]+E_{x\sim p_{g}}[\log (1-D_{G}^{*}(x))]
\end{equation}
then $p_{g}$ converges to $p_{data}$
\end{proposition}

\noindent In practice, adversarial nets represent a limited family of $p_{g}$ distributions via the function $G(z; \theta_{g})$ and we optimize $\theta_{g}$ rather than $p_{g}$itself.

Similar to Fridirich's $\epsilon$ \text{-} security steganography, we define a $\epsilon$ \text{-} security for generative cipher system based on Jensen\text{-}Shannon divergence:
\begin{equation}
D_{JS}(p_{cipher}, p_{data}) \leq \epsilon
\end{equation}

\noindent Ideally, when the generator is optimal, i.e., $\epsilon = 0$, the system can be considered perfectly safe to statistical analysis in steganalysis. In generative cipher, we not only require the $D_{JS}(p_{g}, p_{data})=0$, but we also require the generator to satisfy the request for message extraction as shown in PPC (2).

\subsection{A Generative Perspective}
Before the modern era, classical cryptography focused on message confidentiality (i.e., encryption)¡ªconversion of messages from a comprehensible form into an incomprehensible one and back again at the other end, rendering it unreadable by interceptors or eavesdroppers without secret knowledge (namely the key needed for decryption of that message). This procedure can be formalized as follow:
\begin{equation}
Dec(Enc(m,k),k) = m
\end{equation}
\begin{equation}
Enc(m,k)=\mathop{\argmin}_{c \sim p_{cipher}}{D_{JS}(p_{cipher},p_{uniform})}
\end{equation}
\begin{equation}
Dec(c,k)=m   
\end{equation}
\noindent where $Enc(.)$ is a encryption operation. Security of the key used should alone be sufficient for a good cipher to maintain confidentiality under an attack. This representation can be considered as a constrained cipher generation problem. $D_{JS}(p_{cipher},p_{uniform})$ is the Jensen\text{-}Shannon divergence between the cipher's distribution and the uniform distribution. Note that, the only difference from perfect cipher scheme is that it is not intended to fulfill PCC(3). Ciphertexts produced by a classical cryptography (and some modern ciphers) will reveal statistical information about the plaintext, and that information can often be used to break the cipher. The aim of classical cryptography is to tend to flatten the frequency distribution to the uniform distribution.

Currently, the state-of-the-art methods of cover modification steganography can be viewed as a constrained coding problem, which minimizing the distortion between cover and stego with Syndrome Trellis Coding (STC)\cite{ref_article_Filler}. The embedding and extraction mappings are realized using a binary linear code $C$ :
\begin{equation}
Ext(Emb(x,m)) = m
\end{equation}
\begin{equation}
Emb(x,m)=\mathop{\argmin}_{y \in C(m)}{D(x,y)}
\end{equation}
\begin{equation}
Ext(y)= Hy
\end{equation}
\noindent where $x$ is cover, $m$ denotes message, $y$ is stego. $D(x, y)$ is the distortion function. $Emb(.)$ and $Ext(.)$ denotes embedding and extraction operation which also can be considered as encryption and decryption operations. Embedding processing is an optimum problem to find a stego $y$ that satisfying the message extraction condition and minimizing the distortion, simultaneously. The embedding problem can be optimally solved by the Viterbi algorithm. This implementations of steganography that lack a shared secret are forms of security through obscurity which is the reliance on the secrecy of the design or implementation as the main method of providing security for a system or component of a system. Furthermore, although stego y is highly correlated with specific cover $x$, a well-trained classifier that training on data set $X$ and $Y$ is able to perform steganalysis.

Similarly, in this paper, we give a representation of optimization problem for generative cipher:
\begin{equation}
Dec(Gen(m,k),k) = m
\end{equation}
\begin{equation}
Gen(m,k)=\mathop{\argmin}_{c \sim p_{cipher}}{D_{JS}(p_{cipher},p_{data})}
\end{equation}
\begin{equation}
Dec(c,k)=m=C_{k}c
\end{equation}

\noindent where $Gen(.)$ is a generator.$Dec(.)$ denotes the decryption operation. $C_{k}$ is an extract matrix based on the secret key $k$. This representation can be considered as a constrained cipher generation problem. $D_{JS}(p_{cipher},p_{data})$ is the Jensen-Shannon divergence between the model's distribution and the data. Note that our generative steganography is a key-dependent steganographic scheme adhere to Kerckhoffs's principle. We will give the details of the $C_{k}$ with a practical algorithm in the next section. It is important to note that an explicit cover $x$ is unnecessary. Stego $y$ does not depend on any specific cover, it is regarded as sampling from generator distribution $p_{g}$.The emergence of the generative adversarial network makes the generative cipher scheme will be more and more attention to how to ensure the accuracy of information extraction.

More specifically, we also formulate the procedure of finding ciphertext $y$ as an optimization problem. Let $m$ be the message and $k$ be the secret key shared by two parties. Using this notation we define the ¡°closest¡± encoding $\hat{z}$ via:
\begin{equation}
\hat{z} = \mathop{\argmin}_{z}{L_{m}(z|m,k)+L_{p}(z)}
\end{equation}

\noindent where $L_{m}$ denotes the message loss, which constrains the generated ciphertext given the message $m$ and the extract key $k$, $L_{p}$ denotes the prior loss, which penalizes unrealistic ciphertext. We get generative cipher $y = G(\hat{z})$. The details of the proposed loss function will be discussed with a practical generative cipher in the section IV.

\begin{figure}[!tb]
\begin{minipage}[t]{0.5\linewidth}
\centering
\includegraphics[width=2.3in]{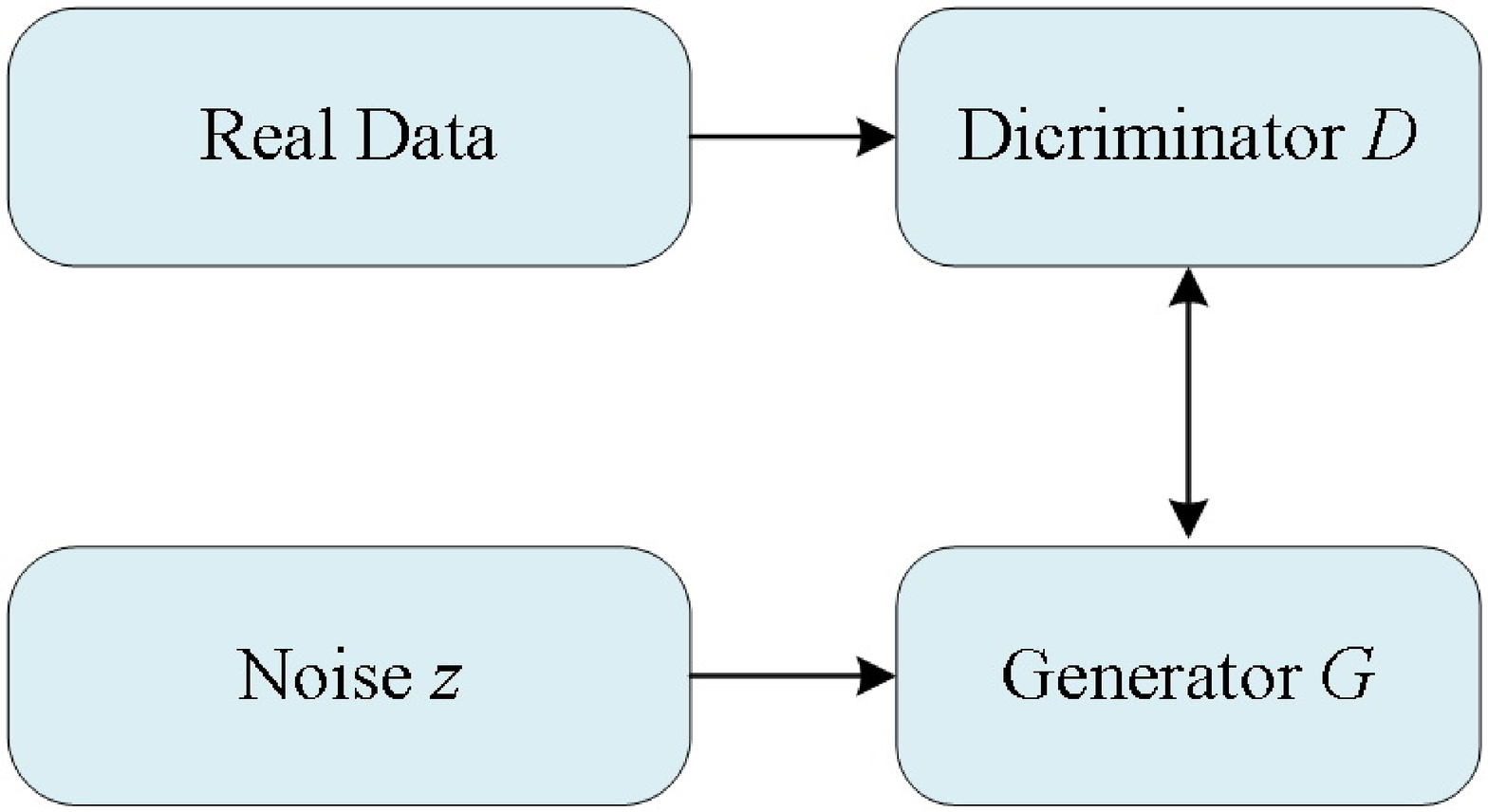}
\quad
{(a) Training a generator with GAN.}
\end{minipage}
\begin{minipage}[t]{0.5\linewidth}
\centering
\includegraphics[width=2.3in]{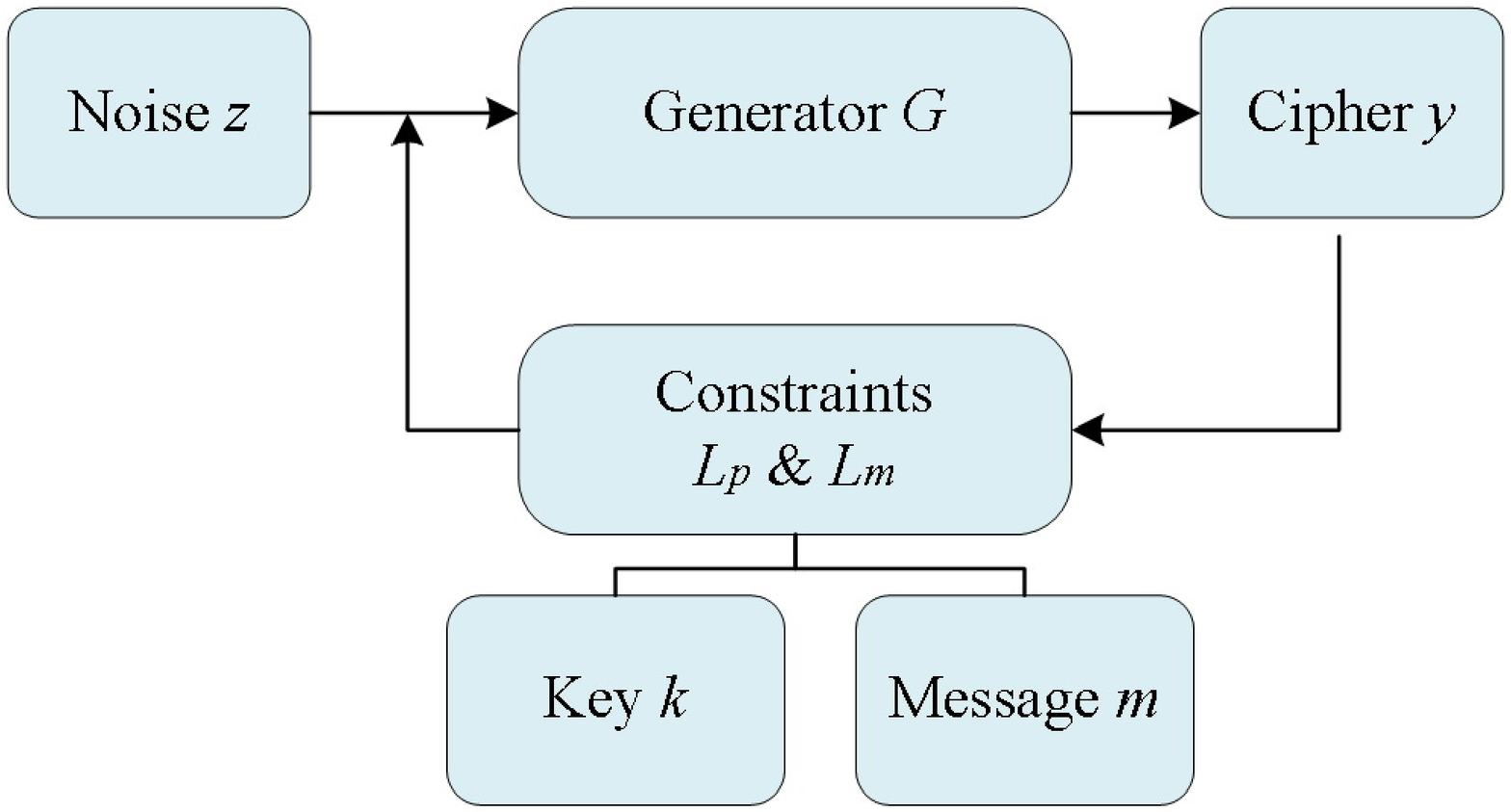}
\quad
{(b) Sampling a cipher with constraints.}
\end{minipage}
\caption{Constrained Cipher Generation.}
\label{fig_constraint}
\end{figure}

In our design, in order to minimize the loss function, we first train a generator, as shown in Fig.~\ref{fig_constraint}(a). As section 3.1 discussed, in this paper generator is constructed by GAN, the training target of GAN is to reach the optimum state of $C(G)$,In which $C^{\star}= - log(4)$ is the global minimum of $C(G)$ in the proposed scheme. Ideally, it indicates $p_{stego} = p_{data}$.

Next, we keep generator fixed to sampling cipher, as shown in Fig.~\ref{fig_constraint}(b). Back-propagation to the input data is introduced to optimize the coding of the input data $z$ on GAN. The back-propagation based methods require specifically designed loss function. In this task, we use $L1$ distance as $L_{m}$.
\begin{equation}
z \leftarrow z - \eta_{z}\nabla_{z}L
\end{equation}
\begin{equation}
\nabla_{z}L = - \frac{\partial{(L_{m}+ L_{p})}}{\partial{z}}
\end{equation}
Similar to ~\cite{ref_article_Yeh}, we iteratively update $z$ using back-propagation by Eq. (23)-(24). After enough training iterations, the input data $z$ on GAN would get optimized to make the loss minimum.

\subsection{A Toy Example for Visual Representation}
\begin{figure}[!htb]
\begin{minipage}[t]{0.5\linewidth}
\centering
\includegraphics[width=2.3in]{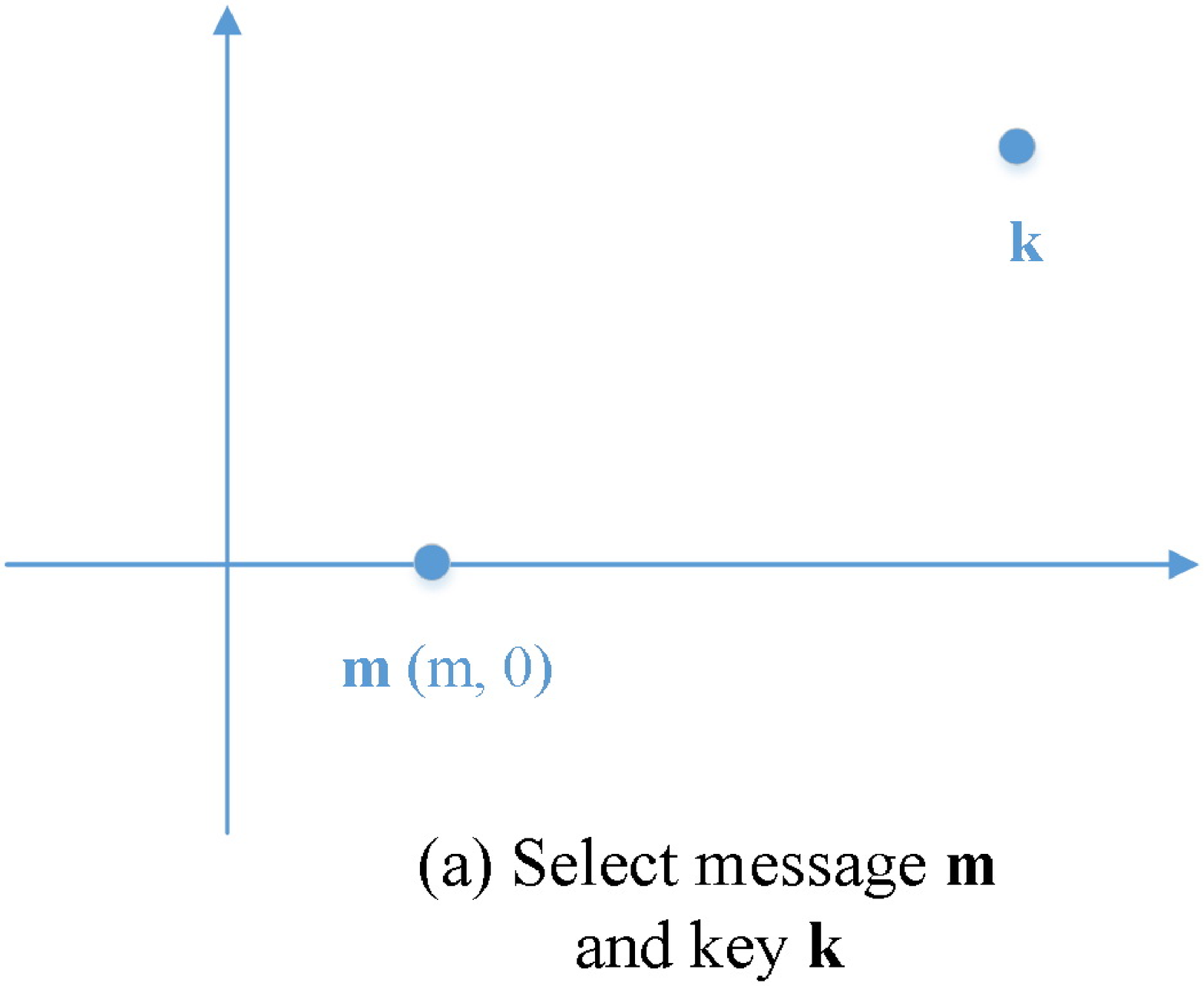}
\end{minipage}%
\begin{minipage}[t]{0.5\linewidth}
\centering
\includegraphics[width=2.3in]{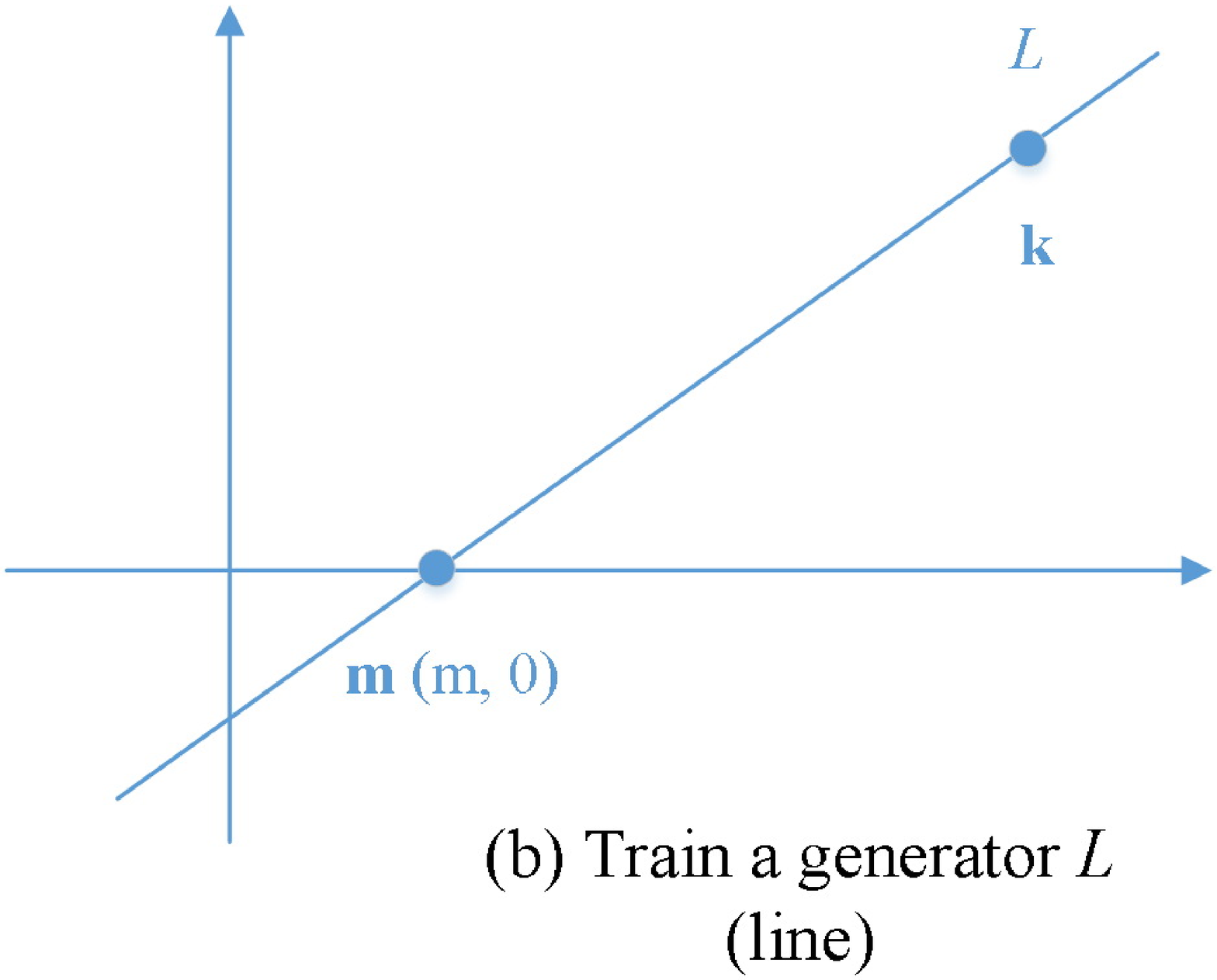}
\end{minipage}
\quad
\begin{minipage}[t]{0.5\linewidth}
\includegraphics[width=2.4in]{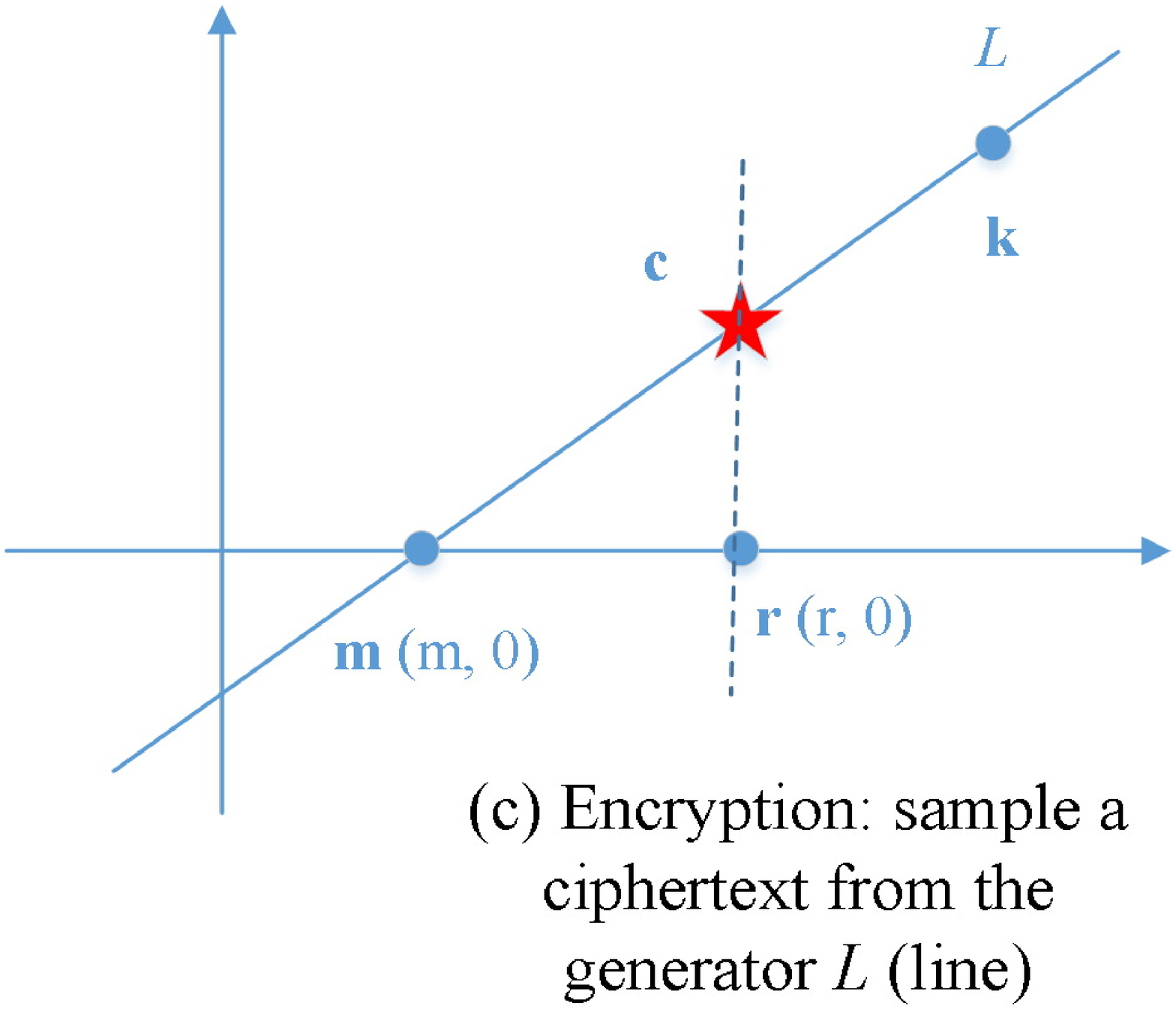}
\end{minipage}
\begin{minipage}[t]{0.5\linewidth}
\includegraphics[width=2.3in]{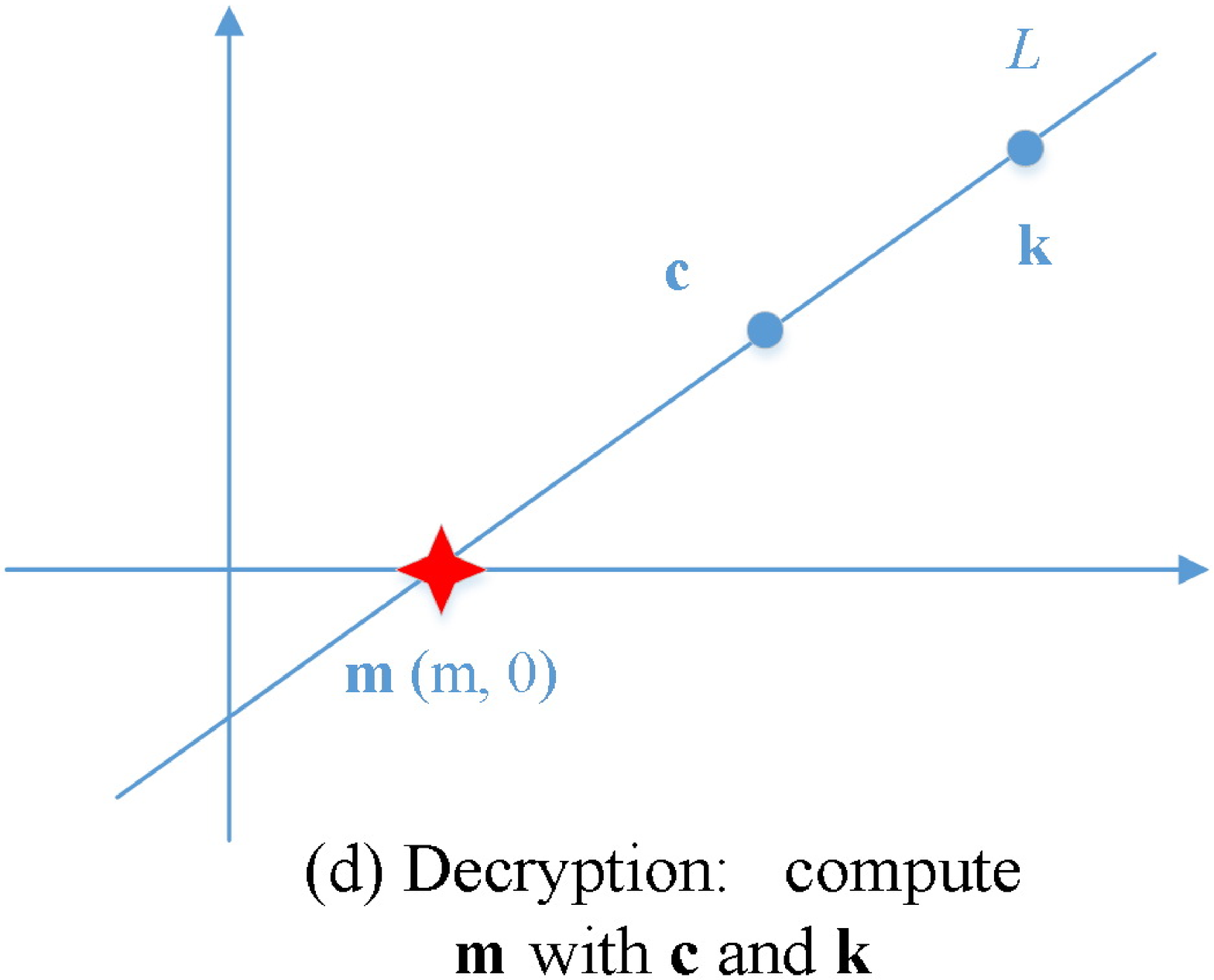}
\end{minipage}
\quad
\begin{minipage}[t]{0.5\linewidth}
\includegraphics[width=2.3in]{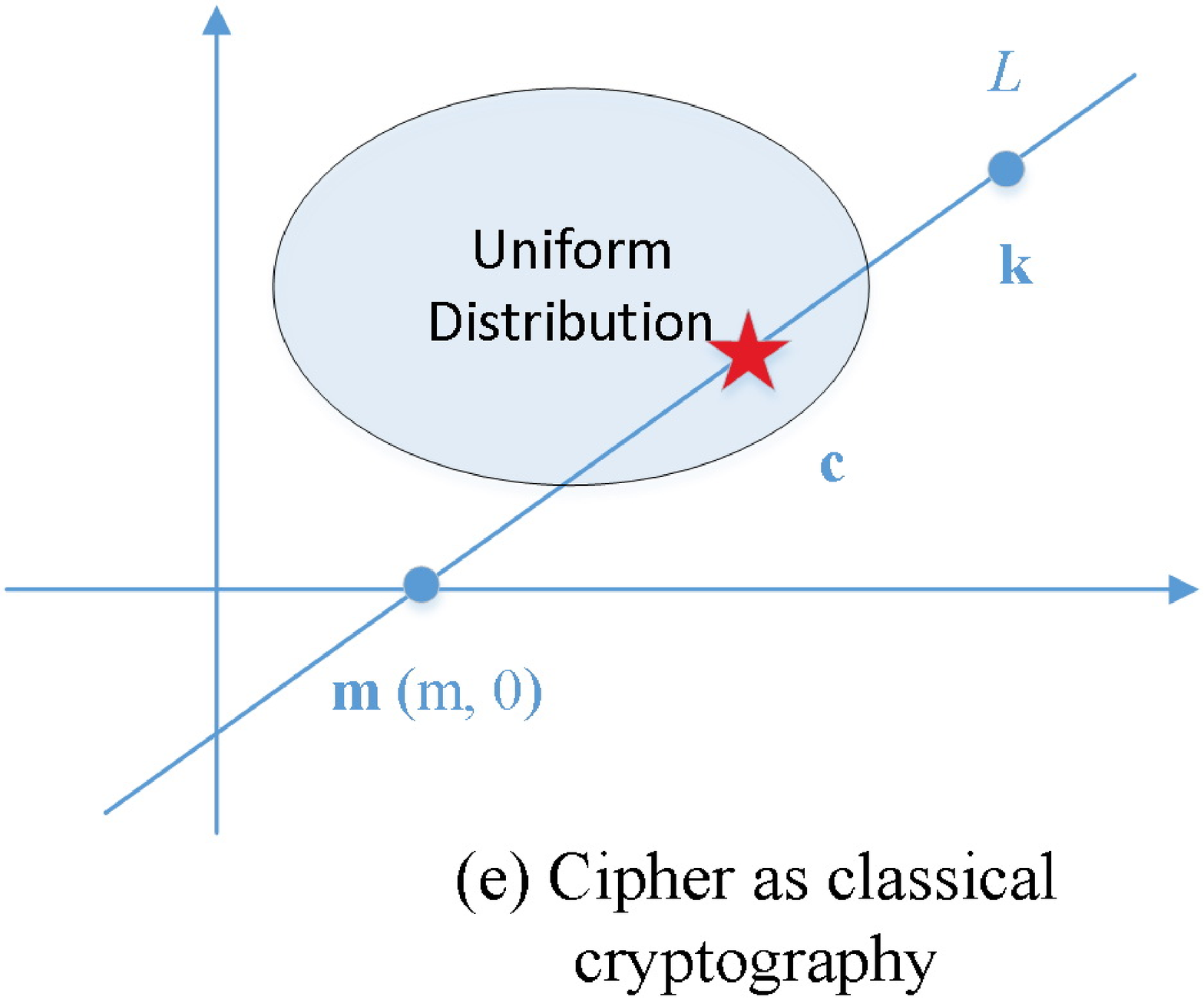}
\end{minipage}
\begin{minipage}[t]{0.5\linewidth}
\includegraphics[width=2.3in]{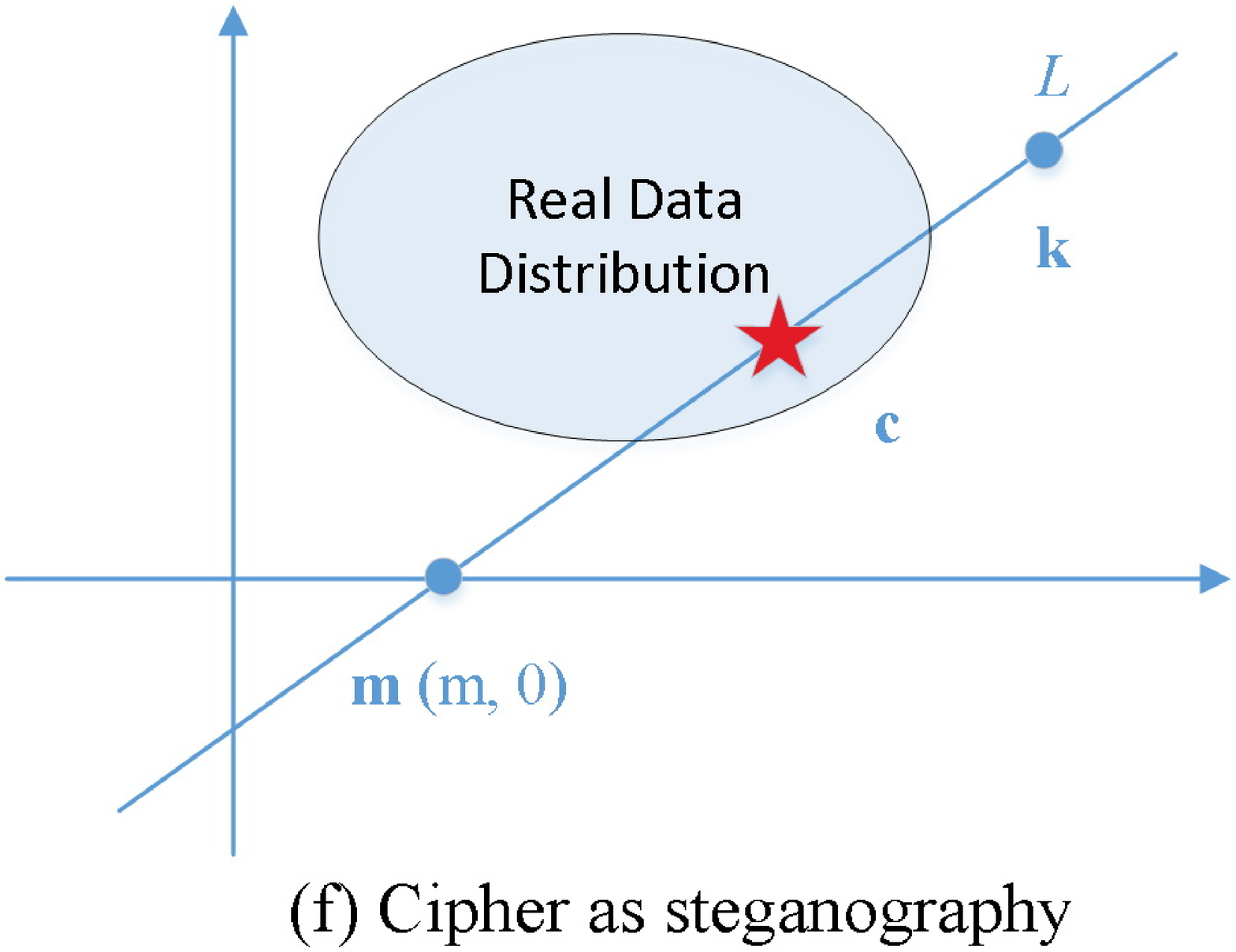}
\end{minipage}
\caption{A Toy Example.}
\label{fig_toy}
\end{figure}

In order to make the difference between stegonography and classical cryptography more clearly. In this subsection, we proposed a simple toy cipher with a simple line on the 2D plane.
Suppose that a coordinate point $(m,0)$ on the X-axis in the plane coordinate represents a secret information $m$. The Shared key $k(k_{x},k_{y})$ may be any point on the plane except for points on the X-axis, as shown in Fig.~\ref{fig_toy}(a). Two points $m$ and $k$ define a unique straight line $L(m, k)$ as shown in Fig.~\ref{fig_toy}(b). In this case, $L$ can be considered as the simplest generator. The sender selects a random number, $r$, then sample a point $c(c_{x},c_{y})$ according to the line, which can be regarded as the corresponding ciphertext, as shown in Fig.~\ref{fig_toy}(c). It's easy for receiver to restore the $m$ by intersection of the $L(c, k)$ and the X-axis as shown in Fig.~\ref{fig_toy}(d). If the ciphertext follow a uniform distribution, this cipher (encryption algorithm) is a classical cryptography shown in Fig.~\ref{fig_toy}(e). If the ciphertext follow a real data distribution, this cipher (steganography algorithm) is a generative steganography, as shown in Fig.~\ref{fig_toy}(f).


In this simple encryption scheme above, this scheme can only resist low-level Ciphtext-only attack. While the attacker has no channel providing access to the plaintext prior to encryption, in all practical ciphertext-only attacks, the attacker still has some knowledge of the plaintext. Cryptographers developed statistical techniques for attacking ciphertext, such as frequency analysis. Every modern cipher attempts to provide protection against ciphertext-only attacks. In the next section, we will present a practical, generated cipher solution with a powerful generator.

\section{Digital Cardan Grille}
A simple practical data-driven cipher method called Digital Cardan Grille is proposed using semantic image inpainting. To fill missing regions in images, our method for cipher generation utilizes the generator $G$ and the discriminator $D$, both of which are trained with uncorrupted data. After training, the generator $G$ is able to take a point $z$ drawn from $p_{Z}$ and generate an image mimicking samples from $p_{data}$. Before constructing the practical generative cipher or steganography algorithm, we hypothesize that a generator has already met $D_{JS}(p_{cipher}, p_{data})=0$, and then we can focus on how to design a scheme to ensure that messages extracted correctly. In this paper, the message is written to the uncorrupted region that needs to be keep in the corrupted image, the stability of the message was guaranteed by the generator, the generator stop updating until the ciphertext (stego image) is natural enough.
\begin{figure}[!htb]
\includegraphics[width=\textwidth]{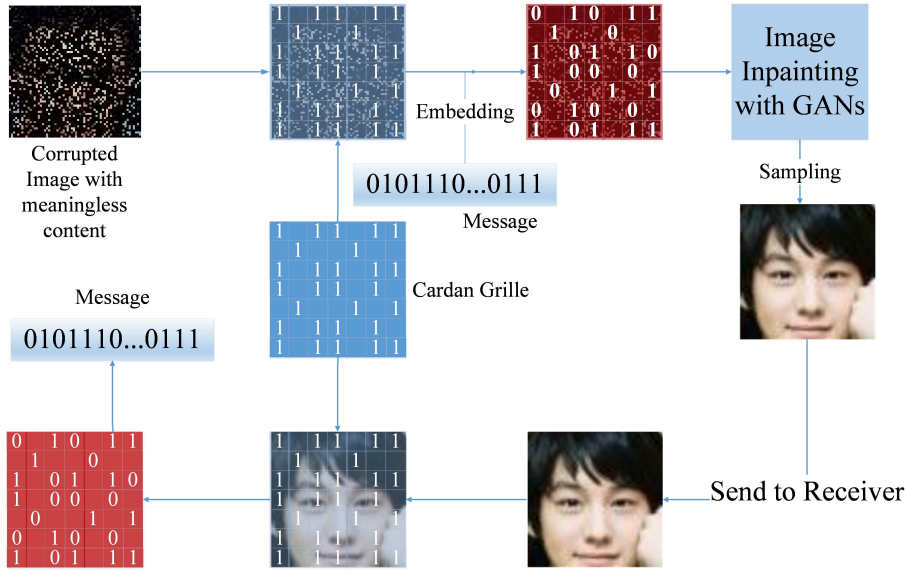}
\caption{The proposed method with Cardan grille.} \label{fig_framework}
\end{figure}
In our framework, as illustrated in Fig.~\ref{fig_framework}, the procedure of cipher is in line with the basic idea of traditional Cardan grille. The sender defines a mask, called Digital Cardan grille, to determine where the message is hidden, and the secret messages go directly to these uncorrupted locations of the input image. Then, an image inpainting method based on GANs is used to finish the image completion. A well-filled image is transmitted to the recipient through the public channel. The receiver extracts a secret message using the Cardan grille shared by the two parties in the reconstructed image. The core of this method is to define generator that not only ensure the consistency of the secret messages but also the natural reality of the ciphertext.

\subsection{Message Preprocessing}
\begin{figure}
\includegraphics[width=\textwidth]{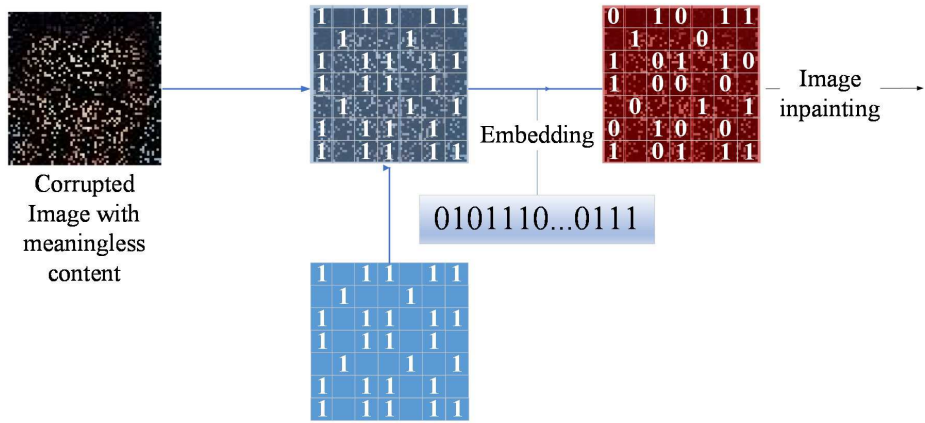}
\caption{Flowchart of Message preprocessing.} \label{fig_messagepreprocessing}
\end{figure}

In this paper, the process of generating ciphertext is decomposed into two steps to simplify the designing, as shown in Fig.~\ref{fig_messagepreprocessing}. First of all, we define a operation $Exn(.)$ for message expansion:
\begin{equation}
m'=Exn(m)
\end{equation}
The secret message $m$ by Cardan grille which is shared by both parties to a new expanded message $m'$ subject to follow constrain:
\begin{equation}
m=Dec(m',k)
\end{equation}
Then, we can get the ciphertext by:
\begin{equation}
c=Gen(m',k)
\end{equation}
We will show that this simple separation trick makes it easier to build a generator.

Firstly, we select the secret input corrupted image $I_{corrupted}$, message $m$, and Cardan grille $C_{k}$. It's important to note the structure and location of this Cardan grille in the corrupted image are shared by both parties. Assume that the size of the corrupt region is $ a\times b$, where $a=b=64$. Then a Cardan grille with same size is defined as:
\begin{equation}
C_{k}=\left[                 
  \begin{array}{ccc}   
    c_{11} & c_{12} \hdots & c_{1b}\\  
    \vdots & \ddots & \vdots\\
    c_{a1} & c_{a2} \hdots & c_{ab}\\  
  \end{array}
\right]                 
\end{equation}

\noindent where $c_{ij}\in \{0,1\}$,$C_{k}$ is the symmetric-key that shared by both parties. Ideally, the Cardan grill is designed to have a $a \times b$ bit key, key length would coincide with the lower-bound on an algorithm's security. A value of 1 represents the parts of the region we want to hide message and a value of 0 represents the parts of the image we cannot write message. Then the message can be written into the uncorrupted regions of the input image. We get a corrupted image contains secrete message shown as $m'$. Note that $m = m' \odot C_{k}$,  $\odot$ denotes the element-wise product operation. The preprocessing is so important that it will transform the image completion into generative cipher (steganography). In the next subsection, we will give the details for the image completion based on the GANs, which complete the generative cipher procedure.

\subsection{Semantic Inpainting for Cipher Generation}
As mentioned above, the image completion used for cipher should satisfy two objectives, one is the rationality of the complete image content, the other is the stability of the message. In this paper we use the a image inpainting method which proposed by Yeh \cite{ref_article_Yeh} based on a Deep Convolutional Generative Adversarial Network (DCGAN).

A binary mask $M$ is used for completion that has values 0 or 1. A value of 1 represents the parts of the image we want to keep and a value of 0 represents the parts of the image we want to complete. Suppose we've found an image from the generator  for some  that gives a reasonable reconstruction of the missing portions. The completed pixels can be added to the original pixels to create the reconstructed image $I_{reconstructed}$:
\begin{equation}
I_{reconstructed} = M \odot m' + (1 - M) \odot G(z)
\end{equation}

It is important to note that cipher generation (semantic inpainting in this case) is not trying to reconstruct the ground-truth image. The goal is to fill the hole with realistic content while hiding information. Even the ground-truth image is one of many possibilities.

In our cipher generation method, three loss functions are defined for searching $\hat{z}$.

\textbf{Contextual Loss}: To keep the same context as the input image, make sure the known pixel locations in the input image $m'$ are similar to the pixels in $G(z)$. We need to penalize $G(z)$ for not creating a similar image for the pixels that we know about. Formally, we do this by element-wise subtracting the pixels in $m'$ from $G(z)$ and looking at how much they differ:
\begin{equation}
L_{contextual}(z) =  \left| \right|M \odot G(z) - M \odot m' \left| \right|_{1}
\end{equation}

\noindent where $||.||1$ is the L1-norm. In the ideal case, all of the pixels at known locations are the same between $m'$ and $G(z)$. Then $G(z)_{i} - m'_{i}$ = 0 for the known pixels $i$ and thus $L_{contextual} (z) = 0$.

\textbf{Perceptual Loss}: To recover an image that looks real, let's make sure the discriminator is properly convinced that the image looks real. We'll do this with the same criterion used in training the DCGAN:
\begin{equation}
L_{perceptual}(z) = log(1 - D(G(z)))
\end{equation}

Contextual Loss and Perceptual Loss successfully predict semantic information in the missing region and achieve pixel-level photorealism.

\textbf{Message Loss}: The key of using image completion for generative cipher (steganography) is that the messages extracted by the Cardan mask $C_{k}$ should be as stable as possible. The pixel value of the corresponding position of the generated image is equal to the value of the secret message.
\begin{equation}
L_{message}(z) = || C_{k}\odot G(z) - C_{k} \odot m' ||_{1}
\end{equation}

Similar to contextual Loss, all of the pixels at hiding locations are the same between $m'$ and $G(z)$. Then $G(z)_{i} ¨C m'_{i} = 0 $ for the known pixels $i$ and thus $L_{message}(z) = 0$. We're finally ready to find $\hat{z}$ with a combination of the all these losses:
\begin{equation}
L(z) = L_{contextual}(z)+ L_{message}(z) + \lambda L_{perceptual} (z)
\end{equation}
\begin{equation}
\hat{z} = \mathop{\argmin}_{z}{L(z)}
\end{equation}
\noindent where $\lambda$  that controls how import the perceptual loss are relative to the message loss. In a particular case, when $C_{k} = M$. $L_{message}$ is the same as $L_{contextual}$ . It is important to note that, $M$ and $C_{k}$ play a different role in cipher generation, the size and value of $C_{k}$ can be different from $M$. $C_{k}$ is used for message encryption£¬while $M$ for image completion.

In practice, for each 8-bit pixel on each layer of color image, we cannot guarantee that the generator will converge to the model that can successfully satisfying $L_{message}(z) = 0$. Intuitively, we believe that the lower bits are affected by the pixel generation, while the high bit has a higher stability. We define a bit plane index (BPI = 1,..8.) to indicate the location of the layer where the message is located. Where, $BPI=1$ represents the lowest significant bit (LSB), and BPI=8 represents the most significant bit(MSB). The element-wise product  is operated on the bit plane level.

\subsection{Message Extraction}
\begin{figure}
\includegraphics[width=\textwidth]{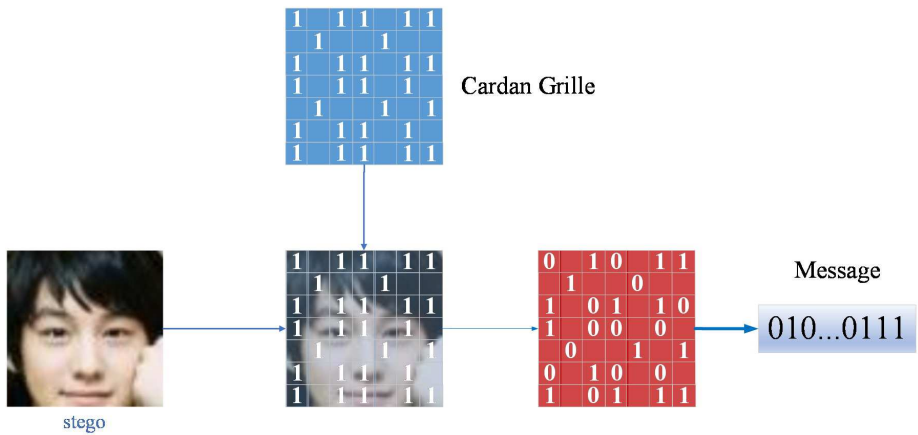}
\caption{Message Extaction using Cardan grille.} \label{fig_messageextract}
\end{figure}

Message extraction for the receiver's is simple as shown in Fig.~ \ref{fig_messageextract}, The receiver will cover the grille directly on the image after reconstruction, and the secret message of the corresponding position can be obtained. The basic operation is as follows:
\begin{equation}
m = I_{reconstructed} \odot C_{k}
\end{equation}

\section{Experiments}
\subsection{Datasets and Settings}
\begin{figure}
\includegraphics[width=\textwidth]{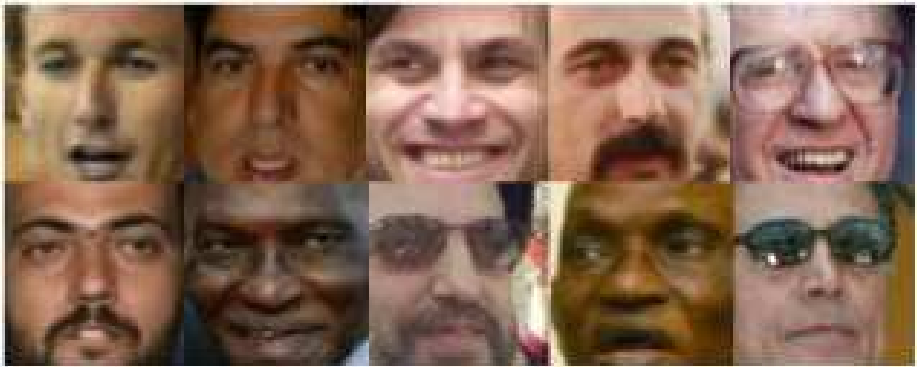}
\caption{Aligned samples form LFW database.} \label{fig_LFW}
\end{figure}

We implemented our adversarial training scheme on the LFW datasets \cite{ref_article_Miller}: a database of face photographs designed for studying the problem of unconstrained face recognition,some samples shown in Fig.~\ref{fig_LFW}. The data set contains more than 13,000 images of faces collected from the web. We use alignment tool to pre-process the images to be $64 \times 64$, as shown in Fig.5. We used the DCGAN model architecture from Yeh et al. \cite{ref_article_Yeh} in this work. I emphasize that we modify Brandon Amos's implementation \cite{ref_article_Amos}. 12000 samples are used for training DCGAN. Our setting of training parameters for image completion is same as the Brandon Amos's. The generative model, G, takes a random 100 dimensional vector drawn from a uniform distribution between [-1; 1] and generates a $64\times64\times3$ image. The discriminator model, $D$, is structured essentially in reverse order. The input layer is an image of dimension $64\times64\times3$, followed by a series of convolution layers where the image dimension is half, and the number of channels is double the size of the previous layer, and the output layer is a two class softmax. For training the DCGAN model, we follow the training procedure in \cite{ref_article_Amos} and \cite{ref_article_Yeh} for optimization. We choose  $\lambda= 0.1$ in all our experiments. In the cipher generation stage, we need to find $\hat{z}$ using back-propagation. We use Adam for optimization and restrict $z$ to [-1; 1] in each iteration, which we observe to produce more stable results. We terminate the back-propagation after 1000 of iterations. We use the identical setting for all testing datasets. The size of grille is fixed as $64 \times 64$ which is same as the size of corrupted image. We intentionally randomize the secret message on all uncorrupted regions so that the stability of embedded messages can be given in a quantitative manner.

We test four random pattern masks different shapes of masks: 1) random pattern masks approximately 20$\%$ missing; 2) 50$\%$ missing masks (randomly horizontal or vertical); 3) 90$\%$ missing complete random masks.

\subsection{Visual Comparison}
\begin{figure}
\includegraphics[width=\textwidth]{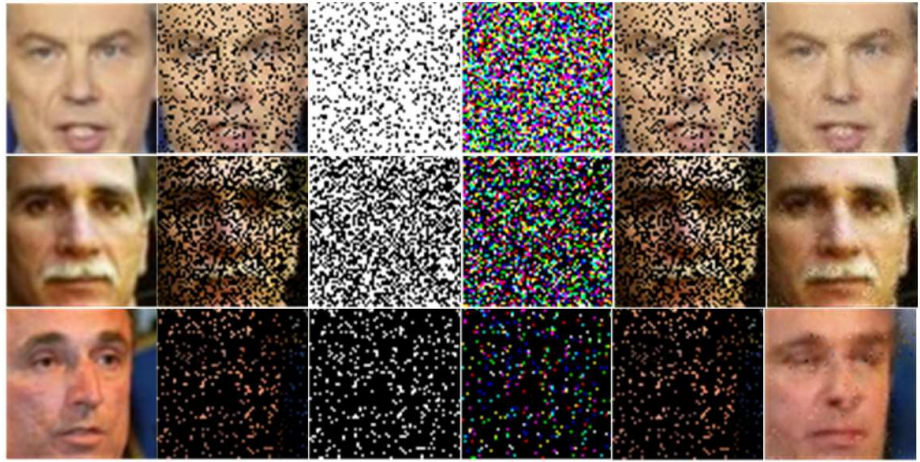}
\caption{Message Extaction using Cardan grille.} \label{fig_wholeprocessdemo}
\end{figure}

Our results are shown in Fig.~\ref{fig_wholeprocessdemo}, which demonstrate that our method can successfully predict the missing content with different random mask. It's important to emphasize that, in our experiment, Cardan grille was randomly generated, and, in all the places that we could write, we wrote the message which is also randomly generated. It is important to note again that cipher generation is not trying to reconstruct the ground-truth image. The goal is to finding a realistic image while encrypt information. Even the ground-truth image is one of many possibilities.

\begin{figure}
\includegraphics[width=\textwidth]{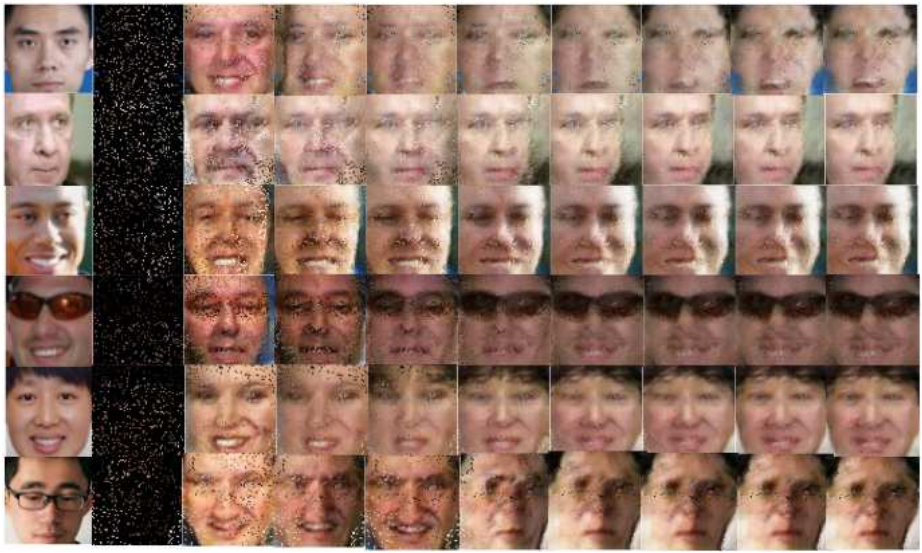}
\caption{For each example, Column 1: Ground-truth image from the dataset. Column 2: Stego corrupted images with random region missing 90$\%$ . Column 3-10 : Samples from the generator as the number of iterations increases.} \label{fig_generationprocess}
\end{figure}

We also show the completion image generation process in Fig.~\ref{fig_generationprocess}, and the number of iterations is from 20 to 2000. We sample 8 generative images form the generator. Note that the we chose some ground-truth images in Fig.~\ref{fig_generationprocess} fall out of LFW database. As can be seen from the Fig.~\ref{fig_generationprocess}, the meaningless serious corrupted image (90$\%$ missing) will be transformed into a sample from pg. In the first few rounds of steps, the visual quality of generator output is low. It can be seen that the complemented image becomes more real as the number of iterations increases.

\begin{figure}[!htb]
\begin{minipage}[t]{0.5\linewidth}
\centering
\includegraphics[width=2.43in]{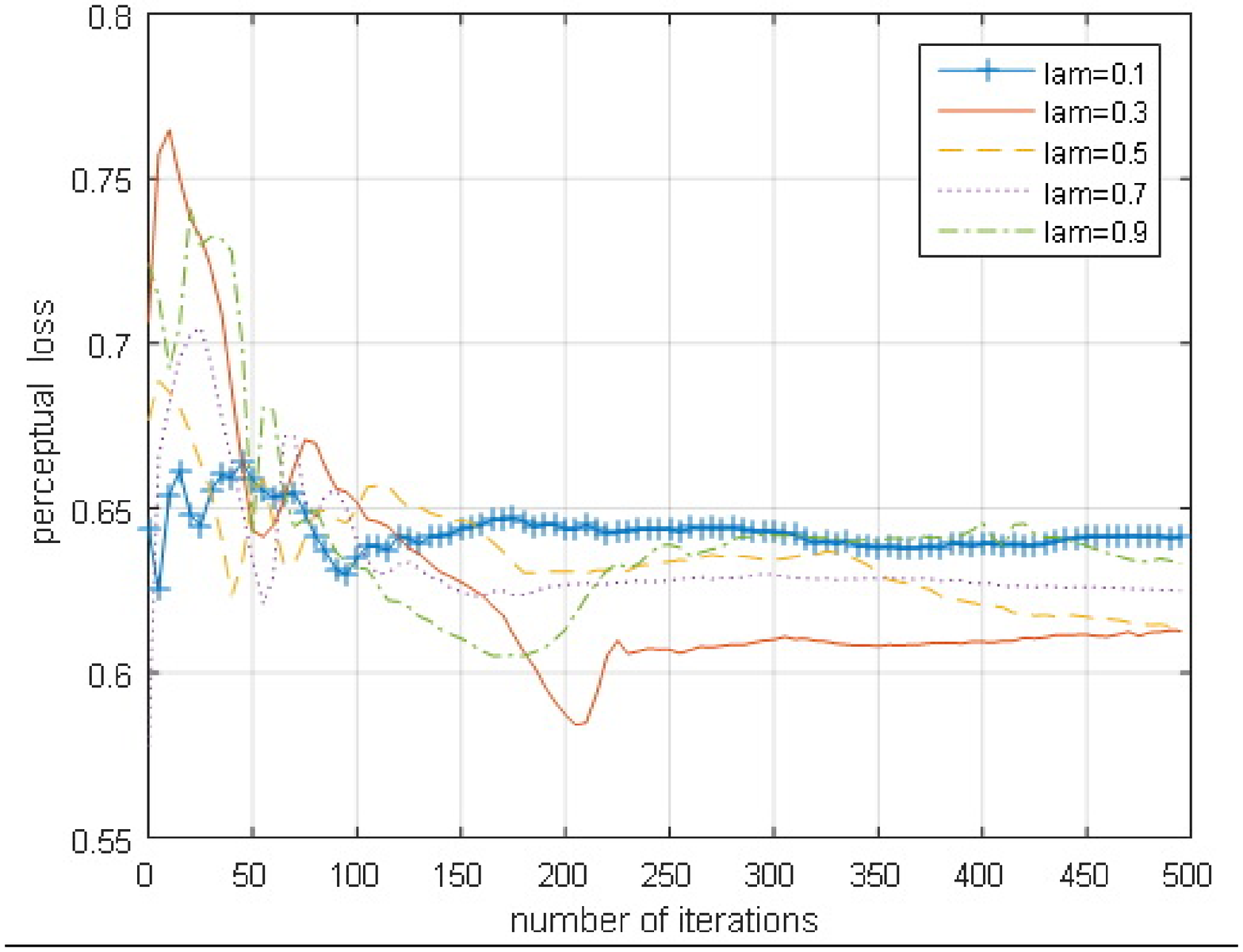}
{(a) Perceptual loss.}
\end{minipage}%
\begin{minipage}[t]{0.5\linewidth}
\centering
\includegraphics[width=2.4in]{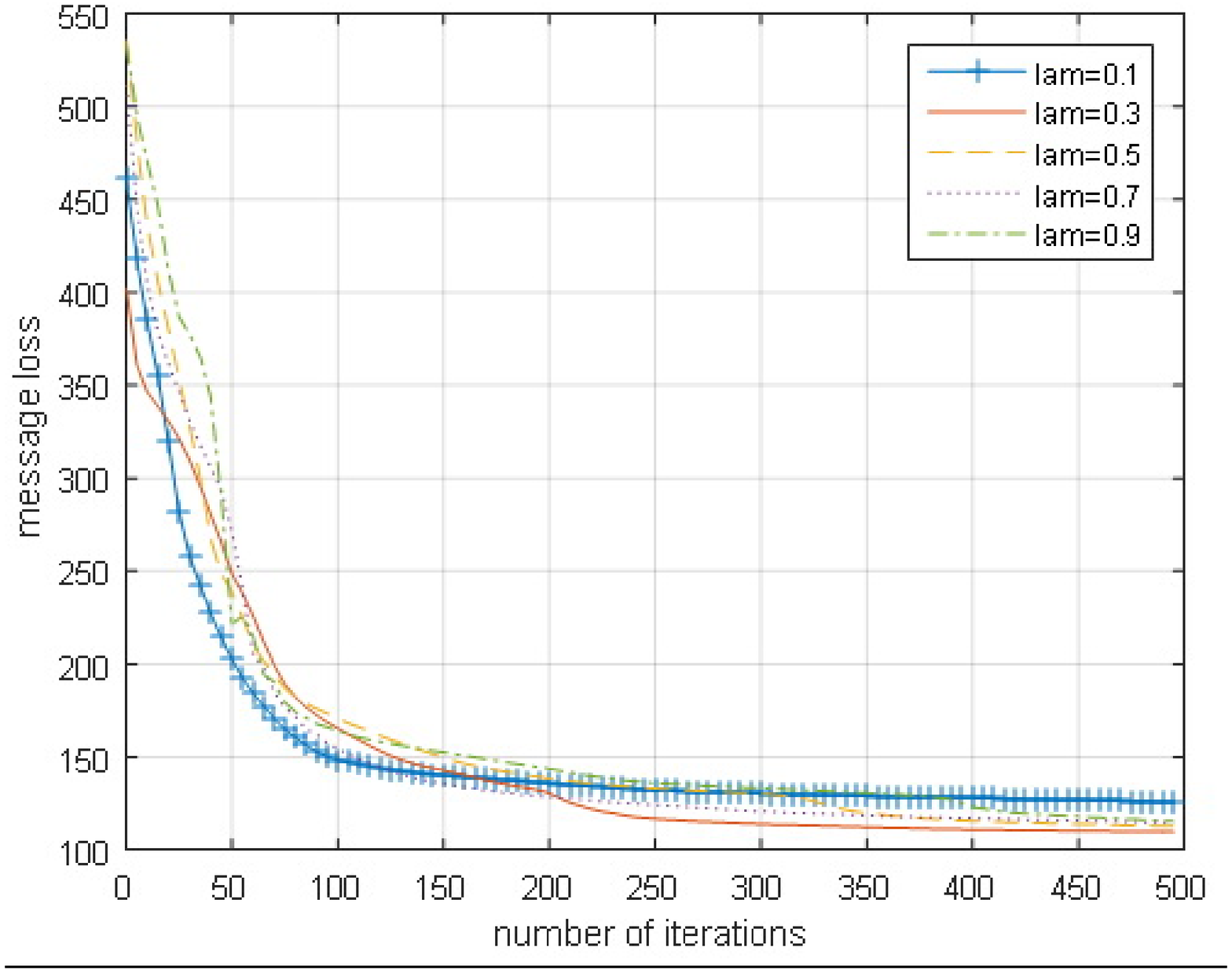}
{(b) Message loss.}
\end{minipage}
\caption{The message loss and perceptual loss.}
\label{fig_loss_perceptual}
\end{figure}

Fig.~\ref{fig_loss_perceptual}(a) and Fig.~\ref{fig_loss_perceptual}(b) shows the message loss and perceptual loss of an image. All images are sampled at 500 iterations from corrupted images with 90$\%$ region missing. In Fig.~\ref{fig_loss_perceptual}(a), in the first few rounds of sampling, the visual quality of output is low, perceptual loss is high. After approximately 250 steps, perceptual loss makes the generated sample more realistic and natural.  In Fig.~\ref{fig_loss_perceptual}{b}, message loss is relatively smooth and stable after 150 steps. This is mainly due to the fact that we keep message loss have more influence on the total loss.

\begin{figure}[!htb]
\includegraphics[width=\textwidth]{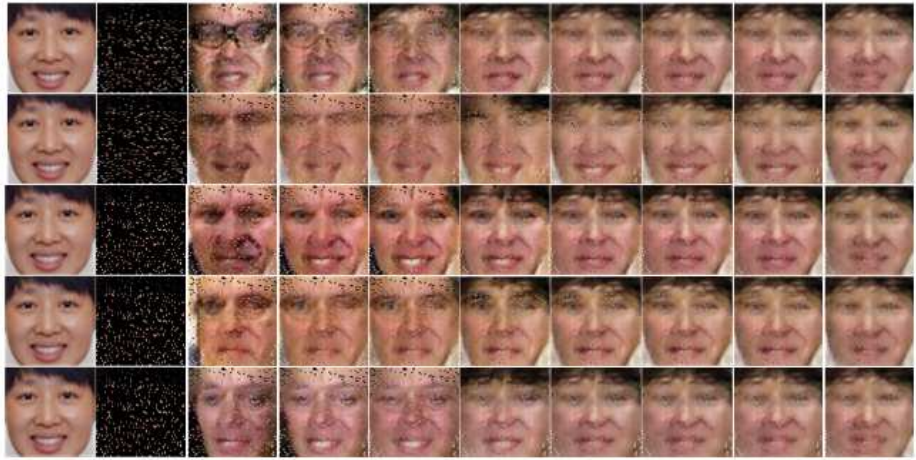}
\caption{Stego Generation for the same image. For each row, Column 1: Ground-truth image. Column 2: Stego corrupted images with random missing 90$\%$. Column 3-10 : Samples as the number of iterations increases.} \label{fig_sameimagecipher}
\end{figure}

We also present the results of the completion of the same image with different $\lambda$ values. It can be seen from Fig.~\ref{fig_sameimagecipher}, although the gap between the generative stego images are large at the beginning, the completion ciphertext images tend to be similar as the number of iterations increases.

\subsection{Quantitative Analysis}
\subsubsection{Efficiency}
All experiments are performed in TensorFlow \cite{ref_article_tensorflow}, on a workstation with a Titan Xp GPU. We take 13.5 hours for training generator with 12000 images. For cipher generation, 1000 iterations for one image take 26s on average. We also denote the encryption blowup factor of our scheme as $l_{EN}$, 1 bit plaintext will be encrypted into $l_{EN}$ bit ciphertext. Take the experiment in Fig.~\ref{fig_sameimagecipher} as an example, $90\%$ region missing means $UncorruptedRate = EmbedingRate = 0.1$, the secret message has a size of equal to $64\times 64\times  UncorruptedRate$ bits, after cipher generation, the corresponding uncompressed ciphertext has a size of $64\times 64\times 8$ bits, the encryption blowup factor is $8\slash EmbedingRate$.

\subsubsection{Decryption-error rate}
\begin{figure}[!htb]
\includegraphics[width=\textwidth]{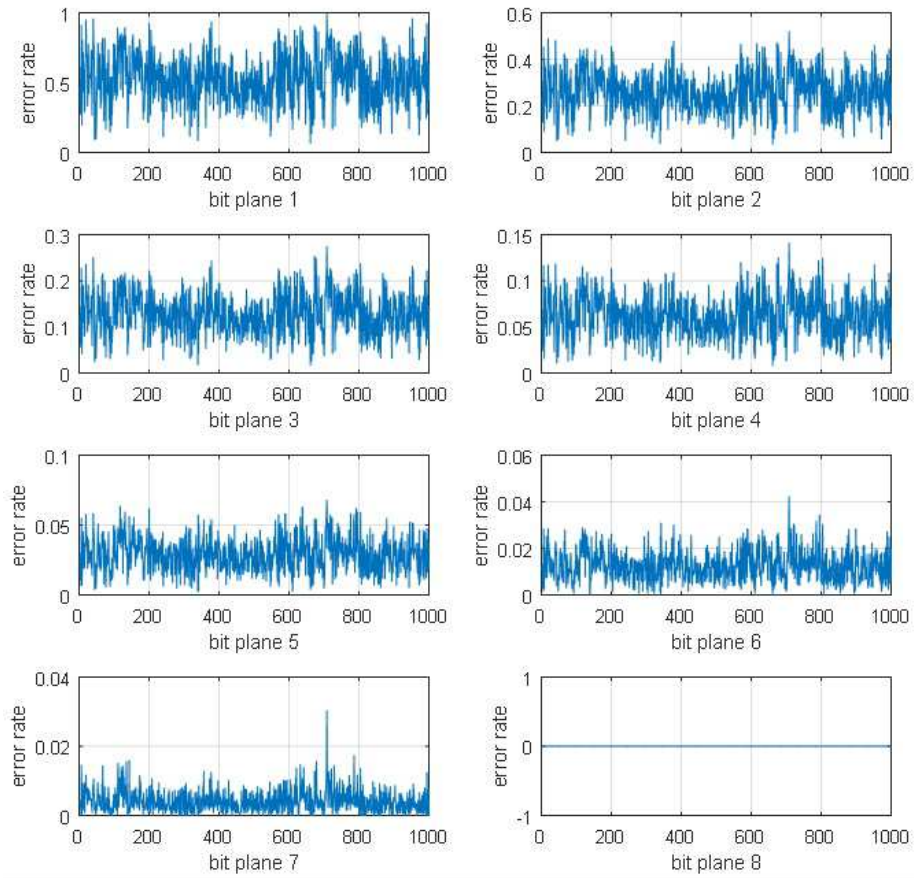}
\caption{Error rate of the message extraction for different bit plane.} \label{fig_messageerrorrate}
\end{figure}

\begin{figure}
\includegraphics[width=3.0in]{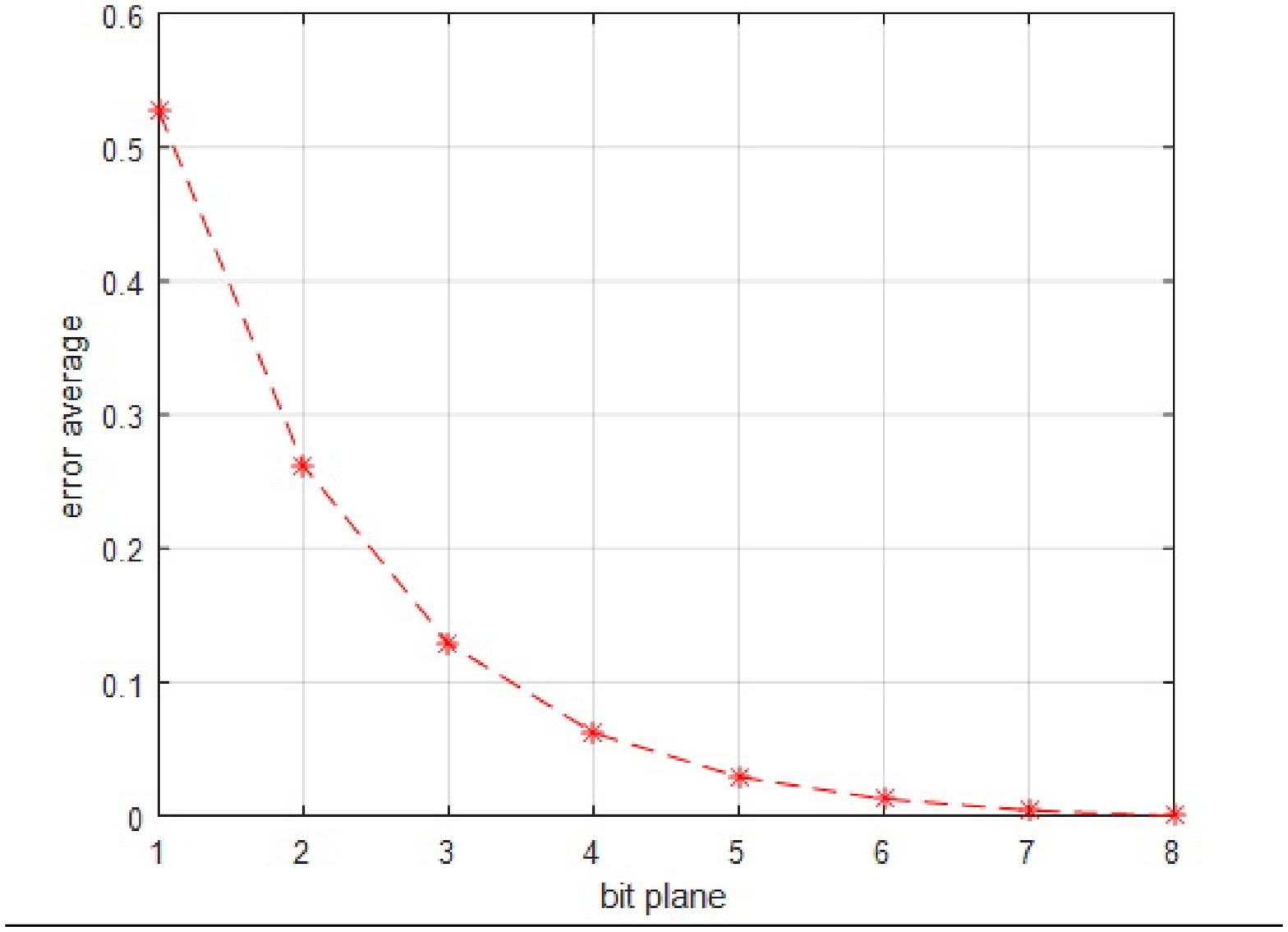}
\centering
\caption{Average Error rate of the message extraction.} \label{fig_messageaverageerrorrate}
\end{figure}

Due to the non-convexity of the models in the training scheme, we cannot guarantee that the generator will converge to the model that can successfully recover the secret message from the steganographic image perfectly. Fig.~\ref{fig_messageerrorrate} shows the relationship between the error rate of the message extraction and the number of iterations with different BPI (1-8). As shown in Fig.~\ref{fig_messageerrorrate}. We do the message embedding and extraction at different BPI for 1000 images which not belonging to the training set. All ciphertext (stego images in this case) are sampled at 3000 iterations from corrupted images with 90$\%$ region missing. As expected, the accuracy of message extraction increased with the increase of BPI. The receiver was able to recover more than 95 $\%$ of messages sent by sender when $BPI\geq3$. Our scheme can perfectly decode the secret encrypted message from the steganographic image at BPI =8. In the Fig.~\ref{fig_messageaverageerrorrate}, the relationship between the average error rate and BPI is given, compare with our work in \cite{ref_article_Jia}, the stability of the message extraction is greatly improved.

\subsubsection{Security for Channel} We steganalyze our digital Cardan grille method using blind steganalyzer for spatial domain and the ensemble classifier. 686-dimensional SPAM features \cite{ref_article_Pevny} and 504-dimensionnal SCRMQ1 features \cite{ref_article_Denemark} with ensemble classifiers \cite{ref_article_Kodovsky} implemented as random forests are used for this experiment. The decision threshold of each base learner is adjusted to minimize the total detection error under equal priors on the training set:
\begin{equation}
P_{E} = \min \frac{1}{2}(P_{FA}+P_{MD}(P_{FA}))
\end{equation}

\noindent where $P_{FA}$, $P_{MD}$ are the probabilities of false alarms and missed detection, respectively.we adjust the threshold to $L/2$ as $P_{E}$ is nowadays considered standard for evaluating the accuracy of steganalyzers in practice.

\begin{figure}[!htb]
\begin{minipage}[t]{0.5\linewidth}
\centering
\includegraphics[width=2.3in]{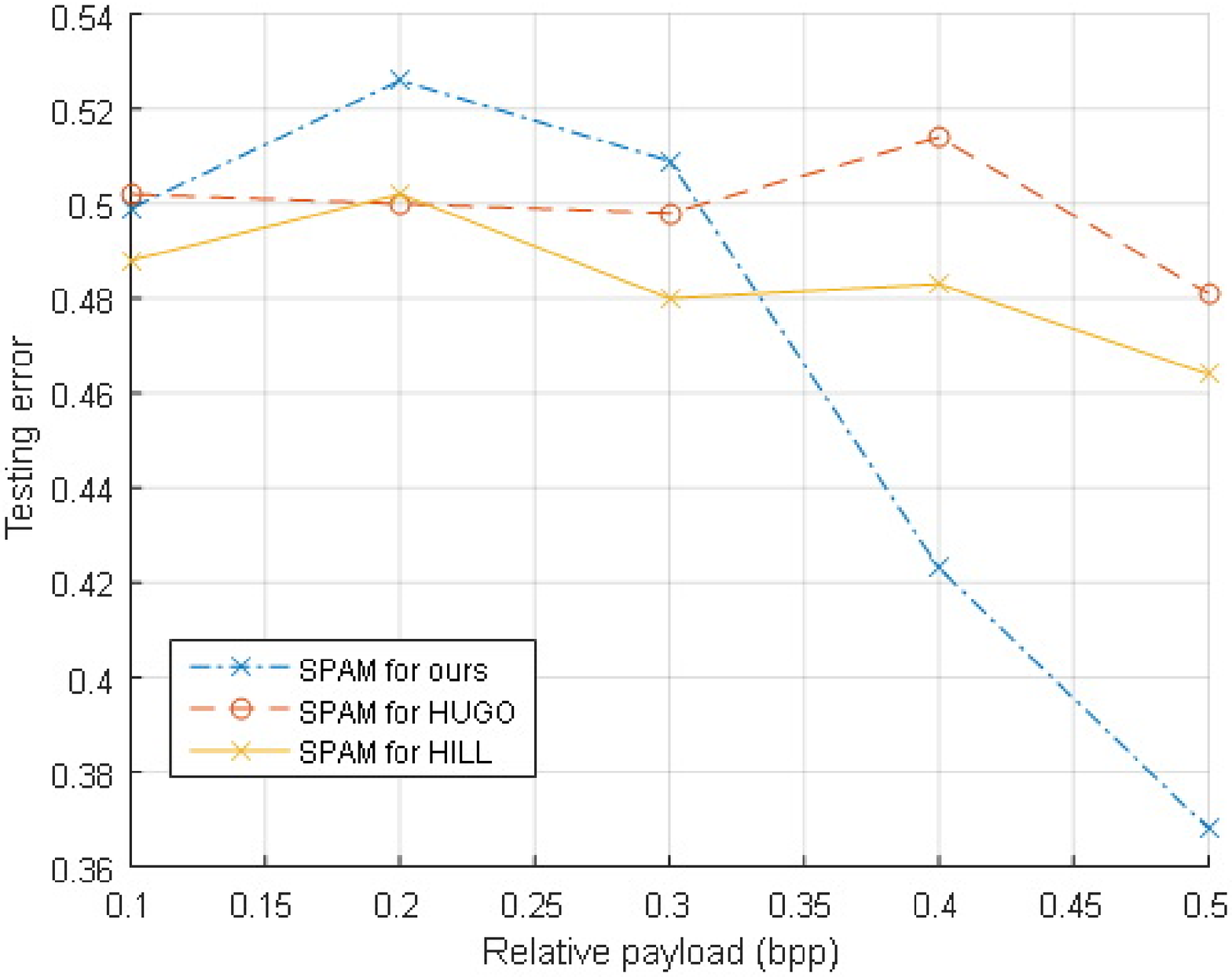}
{(a) SPAM features.}
\end{minipage}%
\begin{minipage}[t]{0.5\linewidth}
\centering
\includegraphics[width=2.4in]{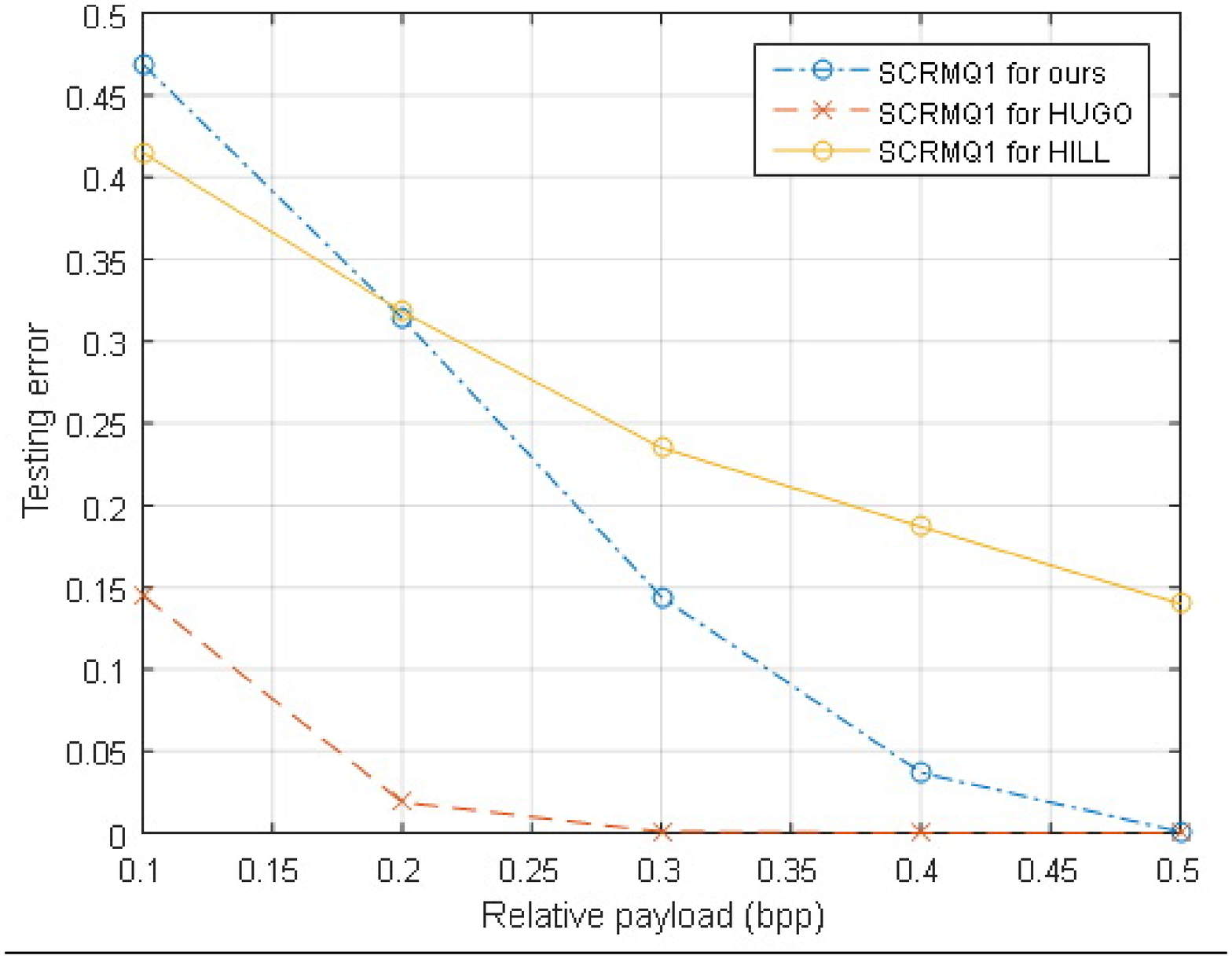}
{(b) SCRMQ1 features.}
\end{minipage}
\caption{Steganalyzer error $P_{E}$ for a ensemble classifier using (a) SPAM features and (b) SCRMQ1 features for five different payloads(0.1 to 0.5 bpp) with HUGO, HILL and our method.}
\label{fig_error PE}
\end{figure}

Different from the traditional steganalyzer for cover modification method, all 1000 stego images and 1000 normal images are generated at 1000 iterations from corrupted images by the image inpainting. The database was divided randomly into two halves, one used for training and the other for testing. The performance is averaged over ten random splits. In Fig.~\ref{fig_error PE}, we plot the progress of the testing error $P_{E}$ as a function of the payloads from 0.1bpp to 0.5bpp (bits per pixel) with BPI = 1 compared with HUGO~\cite{ref_lncs_Pevny} and HILL~\cite{ref_proc_Li} which are considered as advanced steganographic method by minimizing distortion using Syndrome-Trellis Codes.

From the above experiments, it can be seen that the steganography based on sampling can resist the statistical analysis of the steganography, this is mainly due to the fact that, completed stego and normal images can be regarded as samples from the same distribution $p_{g}$. The normal cover and stego does not have a pairwise relationship between the extracted features. As can be seen from the figure, our method has competitive performance with  in the case of low embedding rate.
\begin{figure}[!htb]
\includegraphics[width=3.0in]{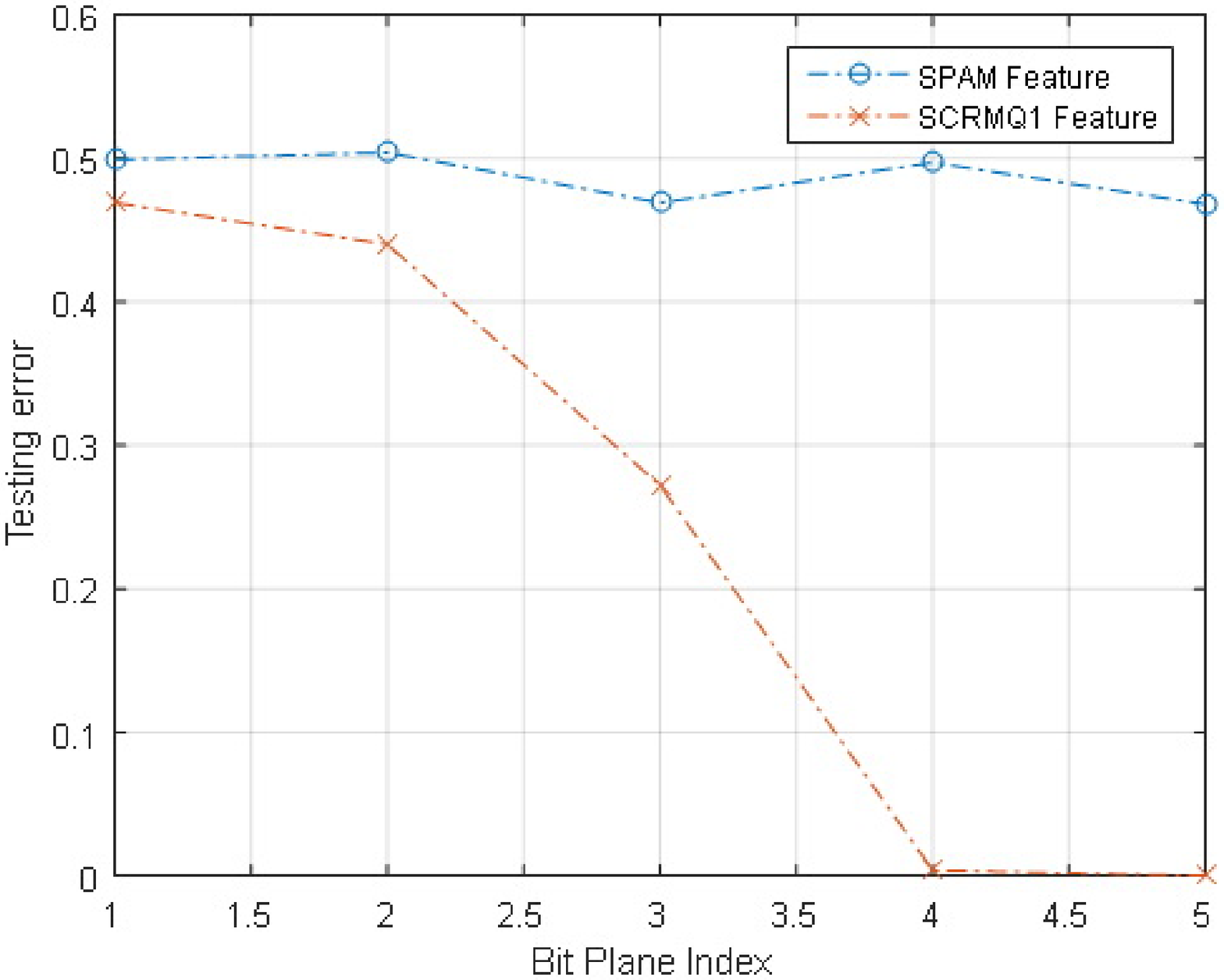}
\centering
\caption{Comparison of methods with five different BPI at 0.1bpp.} \label{fig_dBPI01bpp}
\end{figure}

Fig.\ref{fig_dBPI01bpp} shows the average classification error $P_{E}$ achieved with five different BPI at 0.1bpp with 1000 iterations. As the bit plane index increases, the security of our method decreases on SCRMQ1 feature. SPAM feature does not work. This is mainly because SCRMQ1 is designed for color images, and SPAM is designed for grayscale images.

We also give the error rate for different iterations. This is shown in the Fig.~\ref{fig_numberofiterations} below. After dozens of iterations, SPAM and SCRMQ1 features maintain consistent performance. Experiments show that the resistance to statistical analysis, does not mean that the image generation quality is good enough, in fact, with the increase of the number of iterations, image visual distortion to reduce gradually, as shown in Fig.~\ref{fig_sameimagecipher}.
\begin{figure}[!htb]
\includegraphics[width=3.0in]{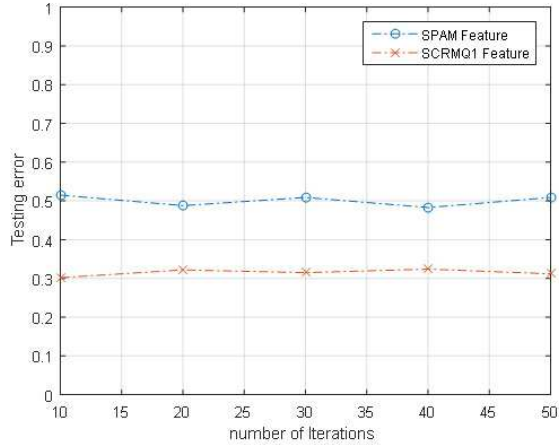}
\centering
\caption{Steganalyzer error $P_{E}$ for a ensemble classifier with different iterations at 0.1bpp with BPI=3.} \label{fig_numberofiterations}
\end{figure}

\subsubsection{Security for Content}
The original Cardan Grille was a literary device for gentlemen's private correspondence. Any suspicion of its use can lead to discoveries of hidden messages where no hidden messages exist at all, thus confusing the cryptanalyst. As in the case of letters/numbers in a random grid, obtaining the grille itself is a chief goal of the attacker. In our method, the size of digital Cardan grille is $M \times N$, where $M$ and $N$ is the size of image. The upper bound key space is $2^{M\times N}$, in this our particular case, key space here is $2^{M\times N \times UncorrouptedRate}$ which means that the message only written on the uncorrupted pixels. But all is not lost if a grille copy can't be obtained. Frequency analysis will show a distribution of cipher. The problem, easily stated though less easily accomplished, is to identify the transposition pattern and so decrypt the ciphertext. Possession of several messages written using the same grille should be avoided. In each communication, the two parties should jointly consume a certain length of the key.

\section{Conclusion and Future Work}
In this paper, a generative cipher is proposed. The relationship between classical cryptography and steganography is given. Ciphertext are sampling from a well-trained generator. Inspired by the idea of Cardan grille, a practical method of generative cipher is proposed by image completion technology. The results of the experiment and the experimental results verify the promising of such simple method. It reduces the sophistication of the steganography design, which allows researchers in other fields can quickly build a cipher system by this framework.

However, the generator in adversarial network is actually in its infancy. In this paper, we use a simple DCGAN to synthesis natural images. We will focus on more powerful generator which automatic synthesis of realistic images to generate more realistic images. The quality of the generated images does not have a quantitative evaluation standard. It is necessary to continually refine the performance of the generator to ensure that the security of generative cipher.

\end{document}